\documentclass[12pt]{article}
\usepackage{epsfig,amsfonts,amsthm}
\usepackage[normalem]{ulem}
\usepackage{amsmath,amssymb}
\usepackage{array}
\usepackage{amsmath}
\usepackage{amsfonts}
\usepackage{amssymb}
\usepackage{subfig}
\usepackage{wrapfig}
\usepackage{graphicx}
\newcommand{\be}{\begin{equation}}
\newcommand{\ee}{\end{equation}}
\newcommand{\bea}{\begin{eqnarray}}
\newcommand{\eea}{\end{eqnarray}}

\renewcommand{\Re}{\mathrm{Re }}
\renewcommand{\Im}{\mathrm{Im }}
\newcommand{\doublet}[2]{ \left( \begin{array}{c}#1 \\ #2 \end{array}\right) }

\usepackage{color}

\def\lsim{\mathrel{\rlap{\lower4pt\hbox{\hskip1pt$\sim$}}
    \raise1pt\hbox{$<$}}}         %less than or approx. symbol
\def\gsim{\mathrel{\rlap{\lower4pt\hbox{\hskip1pt$\sim$}}
    \raise1pt\hbox{$>$}}}         %greater than or approx. symbol

\def\beq{\begin{equation}}
\def\eeq{\end{equation}}
\def\bea{\begin{eqnarray}}
\def\eea{\end{eqnarray}}

\def\<{\left\langle}
\def\>{\right\rangle}

\usepackage[normalem]{ulem}

\def\lsim{\mathrel{\rlap{\lower4pt\hbox{\hskip1pt$\sim$}}
    \raise1pt\hbox{$<$}}}         %less than or approx. symbol
\def\gsim{\mathrel{\rlap{\lower4pt\hbox{\hskip1pt$\sim$}}
    \raise1pt\hbox{$>$}}}         %greater than or approx. symbol

\def\beq{\begin{equation}}
\def\eeq{\end{equation}}
\def\bea{\begin{eqnarray}}
\def\eea{\end{eqnarray}}

\def\<{\left\langle}
\def\>{\right\rangle}

\newcommand{\gev}{\mathrm{\;GeV}} 
\newcommand{\bt}{\begin{tabular}}
\newcommand{\et}{\end{tabular}}

\usepackage{hyperref}    
    
    \usepackage{tikz}
\usetikzlibrary{decorations.pathmorphing,decorations.markings}

\usepackage{graphicx}%Täl saa laitettuu kuvii. Esim. ps filei. Kun haluu laittaa .epsi tai .eps tiedostoja, niin laittaa eka \includegraphics{nimi.eps}. Sit kääntää pdflatex:lla ja kuva tulee. 

\usepackage{color}

\tikzset{
photon/.style={decorate, decoration={snake,amplitude=2pt, segment length=5pt}, draw=black},
particle/.style={draw=black, postaction={decorate}, decoration={markings,mark=at position .5 with {\arrow[draw=black]{>}}}},
antiparticle/.style={draw=black, postaction={decorate}, decoration={markings,mark=at position .5 with {\arrowreversed[draw=black]{>}}}},
gluon/.style={decorate, draw=black, decoration={coil,amplitude=4pt, segment length=5pt}},
goldstone/.style={draw=green,postaction={decorate},decoration={markings,mark=at position .5 with {\arrow[draw=blue]{>}}}}
}

\allowdisplaybreaks[2]
\begin{document}
\bibliographystyle{OurBibTeX}

\title{\hfill ~\\[-30mm]
\begin{footnotesize}
\hspace{80mm}
HIP-2020-3/TH\\
\end{footnotesize}
\vspace{5mm}
\textbf{Collider signatures of dark CP-violation}        
}
%\date{}

\author{\\[-5mm]
A. Cordero-Cid\footnote{E-mail: {\tt adriana.cordero@correo.buap.mx}} $^{1}$,\ 
J. Hern\'andez-S\'anchez\footnote{E-mail: {\tt jaime.hernandez@correo.buap.mx}} $^{1}$,\ 
V. ~Keus\footnote{E-mail: {\tt Venus.Keus@helsinki.fi}} $^{2,3}$,\ \\
%S. ~F.~King\footnote{E-mail: {\tt King@soton.ac.uk}} $^{3}$,\ \\
S. ~Moretti\footnote{E-mail: {\tt S.Moretti@soton.ac.uk}} $^{~3,4}$,\
D. ~Rojas-Ciofalo\footnote{E-mail: {\tt D.Rojas-Ciofalos@soton.ac.uk}} $^{3}$,\  
D. ~Soko\l{}owska \footnote{E-mail: {\tt dsokolowska@iip.ufrn.br}} $^{5}$
\\ \\
\emph{\small $^1$ Instituto de F\'isica and Facultad de Ciencias de la Electr\'onica,}\\  
\emph{\small 
Benem\'erita Universidad Aut\'onoma de Puebla,
Apdo. Postal 542, C.P. 72570 Puebla, M\'exico,}\\
  \emph{\small $^2$ Department of Physics and Helsinki Institute of Physics,}\\
 \emph{\small Gustaf Hallstromin katu 2, FIN-00014 University of Helsinki, Finland}\\
  \emph{\small $^3$ School of Physics and Astronomy, University of Southampton,}\\
  \emph{\small Southampton, SO17 1BJ, United Kingdom}\\
  \emph{\small  $^4$ Particle Physics Department, Rutherford Appleton Laboratory,}\\
 \emph{\small Chilton, Didcot, Oxon OX11 0QX, United Kingdom}\\
  \emph{\small  $^5$International Institute of Physics, Universidade Federal do Rio Grande do Norte,}\\
\emph{\small Campus Universitario, Lagoa Nova, Natal-RN 59078-970, Brazil}\\[4mm]
  }

\maketitle

\vspace*{-10mm}

\begin{abstract}
\noindent
{We study an extension of the Standard Model (SM) in which two copies of the SM-Higgs doublet which do not acquire a vacuum expectation value, and hence are {\it inert}, are added to the scalar sector. The lightest particle from the inert sector, which is protected from decaying to SM particles through the conservation of a $Z_2$ symmetry, is a viable dark matter candidate.
We allow for CP-violation in this extended dark sector and evaluate the $ZZZ$ vertex and its CP-violating form factor in several benchmark scenarios.
We provide  collider signatures of this dark CP-violation in the form of potentially observable asymmetries and cross sections for the $f\bar f\to Z^*\to ZZ$ process at both leptonic and hadronic machines.} 
 \end{abstract}
\thispagestyle{empty}
\vfill
\newpage
\setcounter{page}{1}

\section{Introduction}
\label{intro}

Since the discovery of what looks like the Standard Model (SM) Higgs boson by the ATLAS and CMS experiments at the CERN Large Hadron Collider (LHC) \cite{Aad:2012tfa,Chatrchyan:2012ufa} a great effort has been put into establishing detailed properties of this particle. Although, as of now, all measurements are consistent with the SM predictions, it is possible that this discovered scalar is a part of the larger scalar sector.

There is a number of reasons to believe that the SM of particle physics is not complete. Cosmological observations imply that only around 4\% of the energy budget of the Universe is explained by baryons \cite{Aghanim:2018eyx}. In fact,  85\% of matter in the Universe is often assumed to be in the form of non-baryonic cold, neutral and weakly interacting  Dark Matter (DM) \cite{Jungman:1995df,Bertone:2004pz,Bergstrom:2000pn}, with masses of different proposed candidates varying from a few GeV to a few TeV.

Models with extended scalar sector with a discrete symmetry can provide such a particle, e.g., the Inert Doublet Model (IDM), a 2-Higgs Doublet Model (2HDM) with an unbroken discrete $Z_2$ symmetry \cite{Deshpande:1977rw}. The scalar sector of the IDM contains {1} \textit{inert} doublet, which is $Z_2$-odd, does not develop a Vacuum Expectation Value (VEV) and does not couple to fermions, and {1} \textit{active} $Z_2$-even Higgs doublet, which has a non-zero VEV and couples to fermions in the same way as the SM Higgs doublet, hence also referred to as I(1+1)HDM. The IDM, however constrained, is a viable model that can provide a viable DM candidate (see the latest analyses, e.g., in \cite{Ilnicka:2015jba,Belyaev:2016lok,Belyaev:2018ext,Kalinowski:2018ylg}). However, due to the imposed exact $Z_2$ symmetry, all parameters in the potential are real and there is no room for additional sources of CP-violation. In order to have CP-violation and DM in multi-inert doublet models at least three scalar doublets are needed.

In this work we focus on the I(2+1)HDM: a 3HDM with {2} inert doublets  and {1} active Higgs doublet, where CP-violation appears purely in the inert sector \cite{Keus:2014jha,Keus:2015xya,Keus:2013hya,Cordero-Cid:2016krd,Keus:2019szx,Cordero-Cid:2018man}. The other possibility, i.e., I(1+2)HDM: a 3HDM with {1} inert doublet plus {2} active Higgs doublets has CP-violation in the extended active sector \cite{Grzadkowski:2009bt, Osland:2013sla}. However, this leads to significant restrictions on the amount of CP-violation by SM Higgs data, as the Higgs particle observed at the LHC is very SM-like, and by contributions to the Electric Dipole Moments (EDMs) of electron and neutron \cite{Keus:2015hva,Keus:2017ioh}.

In the I(2+1)HDM the active sector is by construction SM-like\footnote{Tree-level interactions are identical to those of the SM Higgs, with the exception of possible Higgs decays to new states provided they are sufficiently light. At loop level, additional scalar states contribute to Higgs interactions, such as in the $h\to gg, \gamma \gamma $ and $Z\gamma$.}. 
The inert sector contains 6 new scalars, 4 neutral and 2 charged ones. With the introduction of CP-violation in the inert sector, the neutral inert particles will have mixed CP quantum numbers. Note that the {inert} sector is protected by a conserved $Z_2$ symmetry from coupling to the SM particles, therefore, the amount of CP-violation introduced here is not constrained by EDM data. The DM candidate, in this scenario, is the lightest state amongst the CP-mixed inert states which enlivens yet another region of viable DM mass range, with respect to both I(1+1)HDM and CP-conserving I(2+1)HDM \cite{Keus:2014jha,Keus:2015xya,Keus:2013hya,Cordero-Cid:2016krd,Keus:2019szx,Cordero-Cid:2018man}. 

The layout of the remainder of this paper is as follows. In section \ref{scalar-potential}, we present the details of the scalar potential and the theoretical and experimental limits on its parameters as well as our Benchmark Points (BPs). 
We then follow with the implementation and calculations of the $f \bar f \to Z^* \to Z Z$ process in section \ref{ZZZ-section}, where $f$ is a generic fermion. 
In section \ref{asymmetry-section}, we discuss observable asymmetries in lepton and hadron colliders. Finally, 
in section \ref{conclusion}, we conclude and present the outlook for our future studies.

\section{The I(2+1)HDM framework}
\label{scalar-potential}
\subsection{The scalar potential}

As discussed in \cite{Cordero-Cid:2018man}, the scalar sector of the model is composed of three scalar doublets:
\be 
\phi_1= \doublet{$\begin{scriptsize}$ H^+_1 $\end{scriptsize}$}{\frac{H_1+iA_1}{\sqrt{2}}},\quad 
\phi_2= \doublet{$\begin{scriptsize}$ H^+_2 $\end{scriptsize}$}{\frac{H_2+iA_2}{\sqrt{2}}}, \quad 
\phi_3= \doublet{$\begin{scriptsize}$ G^+ $\end{scriptsize}$}{\frac{v+h+iG^0}{\sqrt{2}}}.
\label{explicit-fields}
\ee
We impose a $Z_2$ symmetry on the model under which the fields transform as
\be  
\phi_{1} \to -\phi_1,
\quad \phi_{2} \to -\phi_2, 
\quad \phi_{3} \to \phi_3, 
\quad\textrm{SM} \to \textrm{SM}.
\ee
To keep this symmetry exact, i.e., respected by the vacuum, $\phi_1$ and $\phi_2$ have to be the {\it inert} doublets, $\langle \phi_1 \rangle = \langle \phi_2 \rangle =0$, while $\phi_3$ is the \textit{active} doublet, $\langle \phi_3 \rangle =v/$\begin{scriptsize}$ \sqrt{2} $\end{scriptsize} $ \neq 0$, and plays the role of the SM Higgs doublet. 
Here, $h$ stands for the SM-like Higgs boson and $G^\pm,~ G^0$ are the would-be Goldstone bosons. 

The resulting $Z_2$-symmetric potential has the following form \cite{Keus:2013hya,Ivanov:2011ae}\footnote{We ignore additional $Z_2$-symmetric terms that can be added to the potential, e.g.,
$ 
(\phi_3^\dagger\phi_1)(\phi_2^\dagger\phi_3), 
$
$
(\phi_1^\dagger\phi_2)(\phi_3^\dagger\phi_3), 
$
$
(\phi_1^\dagger\phi_2)(\phi_1^\dagger\phi_1)
$
and
$ 
(\phi_1^\dagger\phi_2)(\phi_2^\dagger\phi_2).
$
as they do not change the phenomenology of the model \cite{Cordero-Cid:2018man}.}:
\bea
\label{V0-3HDM}
V_{\rm 3HDM}&=&V_0+V_{Z_2}, \\
V_0 &=& - \mu^2_{1} (\phi_1^\dagger \phi_1) -\mu^2_2 (\phi_2^\dagger \phi_2) - \mu^2_3(\phi_3^\dagger \phi_3) \nonumber\\
&&+ \lambda_{11} (\phi_1^\dagger \phi_1)^2+ \lambda_{22} (\phi_2^\dagger \phi_2)^2  + \lambda_{33} (\phi_3^\dagger \phi_3)^2 \nonumber\\
&& + \lambda_{12}  (\phi_1^\dagger \phi_1)(\phi_2^\dagger \phi_2)
 + \lambda_{23}  (\phi_2^\dagger \phi_2)(\phi_3^\dagger \phi_3) + \lambda_{31} (\phi_3^\dagger \phi_3)(\phi_1^\dagger \phi_1) \nonumber\\
&& + \lambda'_{12} (\phi_1^\dagger \phi_2)(\phi_2^\dagger \phi_1) 
 + \lambda'_{23} (\phi_2^\dagger \phi_3)(\phi_3^\dagger \phi_2) + \lambda'_{31} (\phi_3^\dagger \phi_1)(\phi_1^\dagger \phi_3),  \nonumber\\
 V_{Z_2} &=& -\mu^2_{12}(\phi_1^\dagger\phi_2)+  \lambda_{1}(\phi_1^\dagger\phi_2)^2 + \lambda_2(\phi_2^\dagger\phi_3)^2 + \lambda_3(\phi_3^\dagger\phi_1)^2  + h.c. \nonumber
\eea
All parameters of $V_0$ are real by construction. The parameters of $V_{Z_2}$ can be complex and therefore it is possible to introduce explicit CP-violation in the model. For the relevant\footnote{The parameter $\lambda_1$ can also take a complex value, however, since it is only relevant for dark particle self-interactions, it does not appear in the discussion above.} complex parameters we use the following notation with explicit CP-violating phases: 
\bea 
\label{notation}
&&\mu^2_{12} = \Re \mu^2_{12} +i  \Im\mu^2_{12} = |\mu^2_{12}| e^{i \theta_{12}},\nonumber\\
&&\lambda_2 = \Re \lambda_2 +i  \Im\lambda_2 = |\lambda_2| e^{i \theta_2},
\\
&&\lambda_3 = \Re \lambda_3 +i  \Im\lambda_3 = |\lambda_3| e^{i \theta_3}.
\nonumber
\eea
Note that there is an additional rotation freedom in the doublet space and one of the phases, e.g., that of $\mu^2_{12}$, is non-physical. Using this redefinition we can set $\theta_{12}$ to zero for simplicity by:
\bea 
\phi_1 \to \phi_1 e^{i \theta_{12}/2} ~ \hspace{2mm} ~
& & ~
|\mu^2_{12}| e^{i \theta_{12}} \to |\mu^2_{12}| ,
\nonumber\\
\phi_2 \to \phi_2 e^{-i \theta_{12}/2} ~ ~
& \Longrightarrow & ~
|\lambda_2| e^{i \theta_2} \to |\lambda_2| e^{i (\theta_2+\theta_{12})},
\\
\phi_3 \to \phi_3 ~ \hspace{13mm} ~
& & ~
|\lambda_3| e^{i \theta_3} \to |\lambda_3| e^{i (\theta_3+\theta_{12})}.
\nonumber
\eea

As motivated in \cite{Cordero-Cid:2018man}, we study the \textit{dark hierarchy} limit where the following relations are imposed on the model:
\be 
\mu^2_1 =n\mu^2_2 , \quad \textrm{Re}\lambda_3=n \textrm{Re}\lambda_2 , \quad \textrm{Im}\lambda_3=n \textrm{Im}\lambda_2 ,\quad \lambda_{31}=n \lambda_{23} ,\quad \lambda'_{31}=n \lambda'_{23}.
\ee
Here we have introduced the dark hierarchy parameter $n$, which can change between $0 \leq n \leq 1$. Boundary values reduce the model to the well-known I(1+1)HDM for $n=0$ (the aforementioned IDM) and to the \textit{dark democracy} case for $n=1$, where interactions with $\phi_3$ are the same for both inert doublets \cite{Keus:2014jha,Keus:2015xya,Cordero-Cid:2016krd}. The case of $n>1$ corresponds to a redefinition of states and does not lead to any different phenomenology. 
In the dark hierarchy limit, the only two relevant complex parameters, $\lambda_2$ and $\lambda_3$, are related through the relations
$|\lambda_3 | = n |\lambda_2|$ and $ \theta_3 = \theta_2$.
The angle $\theta_2$ is therefore the only relevant CP-violating phase and is referred to as $\theta_{\rm CPV}$ throughout the paper.

The parameters of $V_{\rm 3HDM}$ can be divided into the three groups, all having different impact on the phenomenology of the model. The active Higgs sector is like in the SM, where $\mu^2_3$ and  $\lambda_{33}$ are fixed through extremum conditions by the value of the Higgs mass
\be 
m^2_h = 2\mu^2_3 = 2\lambda_{33} v^2 = (125 \textrm{ GeV})^2.
\ee
Self-interaction of dark scalars are governed by inert/dark sector  parameters:
\be
\lambda_1, \lambda_{11},\lambda_{22},\lambda_{12}, \lambda'_{12}.
\ee 
These parameters are only constrained through perturbative unitarity and positivity of $V_{\rm 3HDM}$. Apart from that, they do not play any role in our analysis, as they do not influence the  considered DM and collider phenomenology. 

The remaining parameters, i.e., $\mu^2_{1},\mu^2_{2},\mu^2_{12}, \lambda_{31},\lambda_{23},\lambda'_{31},\lambda'_{23},\lambda_{2}$ and  $\lambda_{3}$, are related to masses of the inert scalars and their couplings with the visible sector, therefore, they will have major influence on the phenomenology of the model. These 9 parameters can in principle be determined by independent masses, mixing angles and couplings, as shown below.

\subsection{Physical scalar states}
\label{minimization}

In the $Z_2$-conserving minimum of the potential, i.e., at the point $(0,0,\frac{v}{\sqrt{2}})$ with
$ v^2= \frac{\mu^2_3}{\lambda_{33}}$, the resulting mass spectrum of the scalar particles is as follows.
\begin{enumerate}
\item 
$Z_2$-even fields from the active doublet with masses:
\bea 
&& m^2_{G^0}= m^2_{G^\pm}=0, \nonumber\\
&& m^2_{h}= 2\mu_3^2 =2\lambda_{33} v^2. 
\eea
(Recall that  $h$ is the SM-like Higgs boson and $G^0,G^\pm$ are the Goldstone fields.)

\item 
$Z_2$-odd charged inert fields, $S_{1}^{\pm}$ and $S_{2}^{\pm}$, from the inert doublets which are the eigenstates of the matrix
\be 
\mathcal{M}_C= \left( \begin{array} {cc}
-n \mu_{2 }^{2} + \frac{n}{2} \lambda_{23} v^{2} 
& -\mu_{12}^{2}  \\
- \mu_{12}^{2}   &  -\mu_{2}^{2} + \frac{1}{2} \lambda_{23} v^{2}  \end{array} \right) ,
\ee
with eigenvalues:
\be 
m^2_{S^\pm_{1,2}}
=
\frac{1}{4} \left((n+1)(-2 \mu_2^2+\lambda_{23}v^2)\; \mp \;\sqrt{16 (\mu_{12}^2)^2+(n-1)^2 \left(\lambda_{23} v^2-2 \mu_2^2\right)^2}\right).
\ee
In terms of gauge states from Eq. (\ref{explicit-fields}) $S^\pm_i$ are defined as:
\be 
\left( \begin{array} {c}
S_1^\pm\\
S_2^\pm \end{array} \right )= 
\left( \begin{array} {cc}
\cos \alpha_c & \sin \alpha_c \\
-\sin \alpha_c & \cos \alpha_c 
\end{array} \right )
\left( \begin{array} {cccc}
H_1^\pm\\
H_2^\pm \end{array} \right )
\quad 
\mbox{with} 
\quad 
\tan2\alpha_c = \frac{2 \mu_{12}^2}{(n-1) (\mu_2^2 - \lambda_{23} v^2/2)}.
\ee 
We require $\pi/2 < \alpha_c < \pi$, so that $m_{S_1^\pm} < m_{S_2^\pm}$. 

\item 
$Z_2$-odd neutral inert fields, $S_1,S_2,S_3,S_4$, which are the eigenstates of the mass-squared matrix in the $(H_1,H_2,A_1,A_2)$ basis:
\be 
\mathcal{M}_N= \frac{1}{4}\left( \begin{array} {cccc}
n \;\Lambda^+_c
& -2\mu^2_{12}
& -n \;\Lambda_s
& 0 
\\[2mm]
-2\mu^2_{12}
& \Lambda^+_c
& 0 
& \Lambda_s 
\\[2mm]
-n \;\Lambda_s
& 0 
&  n \;\Lambda^-_c
& -2\mu^2_{12} 
\\[2mm]
0 
& \Lambda_s
& -2\mu^2_{12}
&  \Lambda^-_c 
\end{array} \right ),
\label{neutral-mass-squared}
\ee
with
\be  
\Lambda_s=  2\lambda_2 \sin \theta_{\rm CPV} v^2
\quad
\mbox{and}
\quad 
\Lambda^\pm_c =-2\mu^2_2 +
(\lambda_{23}+\lambda'_{23} \pm 2 \lambda _2  \cos \theta_{\rm CPV}) v^2.
\ee  
A non-zero $\Lambda_s$ introduces mixing between states of opposite CP-parity, $H_i$ and $A_i$. The CP-conserving limit is restored for $\theta_{\rm CPV} = 0,\pi \Rightarrow \Lambda_s=0$.

We diagonalise the neutral mass-squared matrix numerically, $\mathcal{M}_N^{\rm diag} = R^T \mathcal{M}_N R$, to derive our mass eigenstates, $S_i$, in terms of the gauge eigenstates in Eq. (\ref{explicit-fields}), 
\be
\left( \begin{array} {cccc}
S_1\\
S_2 \\
S_3 \\
S_4  \end{array} \right )= R_{ij}
\left( \begin{array} {cccc}
H_1\\
H_2\\
A_1 \\
A_2  \end{array} \right ).
\ee
We adopt a notation where $m_{S_1} < m_{S_2}  <m_{S_3} <m_{S_4}$, hence choosing $S_1$ as DM candidate. In the remainder of the paper, the notations $S_1$ and DM will be used interchangeably.

\end{enumerate}

\subsection{Constraints on the I(2+1)HDM parameters}
\label{constraints}

In this section, we summarise the latest theoretical and experimental constrains that are applicable to our studies, described in details in \cite{Cordero-Cid:2018man}. We also refer the reader to that paper for the discussion of future prospects of detection of the model at future collider experiments.

All considered BPs agree with the following constraints.

\begin{enumerate}
\item Boundedness-from-below of the potential and positive-definiteness of the Hessian as well as perturbative unitarity limits for the couplings, i.e.,  $\lambda_i\leq\,4\,\pi$.

\item Total decay width of the SM-like Higgs particle, $\Gamma_\text{tot}\,=3.2^{+2.8}_{-2.2}$ MeV \cite{Sirunyan:2019twz}, and Higgs invisible branching ratio, influenced by decay channels into new inert particles (if they are light enough, i.e., $m_{S_i} \leq m_h/2$).

\item Higgs signal strengths, in particular the loop contributions to $h\to \gamma \gamma$ decays mediated by charged inert scalars. In Run II, ATLAS reports $\mu_{\gamma \gamma} = 0.99^{+0.14}_{-0.14}$ \cite{Aaboud:2018xdt} while CMS reports  $\mu_{\gamma \gamma} = 1.18^{+0.17}_{-0.14}$ \cite{Sirunyan:2018ouh}. Our BPs are within $1\sigma$ agreement with ATLAS and $2\sigma$ agreement with CMS results.
\item Gauge bosons widths, where we forbid decays of gauge bosons into inert scalars $W^\pm \to S_i S_j^\pm$ and $Z\to S_i S_j,S_i^+S_j^-$ by enforcing:
\be 
\label{eq:gwgz}
m_{S_i}+m_{S^\pm_i}\,\geq\,m_W^\pm,~~
\,m_{S_i}+m_{S_j}\,\geq\,m_Z,\,~~
2\,m_{S_i^\pm}\,\geq\,m_Z.
\ee

\item Agreement with Electro-Weak (EW) precision tests parameterised through the so-alled oblique parameters $S,T,U$ \cite{Altarelli:1990zd,Peskin:1990zt,Peskin:1991sw,Maksymyk:1993zm}.
\item No light and/or long-lived charged particles: $m_{S^\pm_i} > 70$ GeV ($i=1,2$) \cite{Pierce:2007ut} with lifetime $\tau\,\leq\,10^{-7}\,{\rm s}$  \cite{Heisig:2018kfq}.

\item We check the agreement with searches for new particles at colliders, in particular the LEP 2 searches for supersymmetric particles (chiefly, sneutrinos and sleptons) in di-jet or di-lepton channels, re-interpreted for the IDM in order to exclude the region of masses where the following conditions are simultaneously satisfied \cite{Lundstrom:2008ai} ($i,j=2, ... 4$):
\be 
\label{eq:leprec}
m_{S_i}\,\leq\,100\,\gev,\,~~
m_{S_1}\,\leq\,80\,\gev,\,\, ~~
\Delta m {(S_i,S_1)}\,\geq\,8\,\gev.
\ee
All benchmarks are also in agreement with null-results for additional neutral scalar searches at the LHC\footnote{Following the analysis performed in \cite{Kalinowski:2018ylg} for the I(1+1)HDM, which is a model with similar signatures to those studied in this paper.}. In general, current searches at the LHC for multi-lepton final states with missing transverse energy, which in principle could constrain the model studied here, are not sensitive enough due to a relatively large cut on missing transverse energy used in the experimental analyses. This corresponds to a rather large mass splittings between scalars in the dark sector and therefore reduces the production cross sections below current sensitivity. Benchmarks with smaller mass splittings between scalars have large enough cross section to be produced in abundance even at the current stage of the LHC, however, they require smaller cuts on missing energy in order to be detected. 

\item Relic density constraints in agreement with the latest results from the Planck experiment \cite{Aghanim:2018eyx},
$ \Omega_c\,h^2\,=\,0.120\,\pm\,0.001$.

\item Direct detection of DM particles in accordance to
the latest XENON1T results \cite{Aprile:2018dbl}. In the region of masses we are considering in this paper, indirect detection experiments (e.g., FermiLAT)  do not place any additional constraints upon the parameter space. 
\end{enumerate}

As discussed in section \ref{scalar-potential}, phenomenologically relevant parameters, used  to define our BPs in a forthcoming section, are:
\be 
\label{parameters}
|\mu^2_{12}| , \;\lambda_{23}, \; \lambda'_{23}, \; \mu^2_2, \; \lambda_2, \; \theta_{\rm CPV}, n.
\ee 
which we take as input parameters.

\subsection{Selection of BPs}
\label{selection}
\begin{table}[t!]
\begin{center}
\begin{tabular}{|p{2.8cm}|p{1.8cm}|p{1.8cm}|p{1.8cm}|p{1.8cm}|}
\hline
& Point-A & Point-B & Point-C & Point-D \\
[1mm]
\hline
$n$ 	                      & 0.6	     		& 0.5	     & 0.8	      & 0.6 \\
$\lambda'_{23}$	& $-0.16$			& $-0.145$	 & $-0.295$	 & $-0.169$\\
$\lambda_{23}$	& 0.29			& 0.171	    	 & 0.294	  & 0.26\\
$\lambda_{2}$  	& 0.067 		& 0.013	   	 & 0.0009    & -0.2\\
$\theta_{CPV}$	    & $15\pi/16$	& $7\pi/8$ & $31\pi/32$	  & $8\pi/15$	  \\
$\mu^2_2$	         & $-13800$ & $-15900$ & $-3400$	 & $-25300$\\
$\mu^2_{12}$     	& 5050  & 7950	   & 250	      &  13700  \\
[1mm]
\hline
$m_{S_1}$             & 72.3     & 55.4	& 50.9 & 63.2\\
$m_{S_2}$ 	        & 103.3    &63.2	& 51.7 & 78.0 \\
$m^\pm_{S_1}$ 	& 106.2  & 79.1 & 99.1	& 106.3 \\
$m_{S_3}$ 	         & 129.4  & 144.3	& 58.5	& 185.0\\
$m_{S_4}$ 	         & 155.1  & 148.8 	& 59.4	& 213.1\\
$m^\pm_{S_2}$ 	& 157.5  & 159.2	& 111.1	& 204.3\\
[1mm]
\hline
$g_{ZS_1S_2}=g_{ZS_3S_4}$   	& 0.366	& 0.37	& 0.37	& 0.312\\
$g_{ZS_1S_3}=g_{ZS_2S_4}$   	& 0.0397	& 0.007	& 	0.0025 & 0.185\\
$g_{ZS_1S_4}=g_{ZS_2S_3}$ 	    & 0.0401	& 0.007	& 0.0028	& 0.07\\
[1mm]
\hline
\end{tabular}
\caption{The input and derived parameters of our BPs. The masses are given in GeV.}
\label{table-BPs}
\end{center}
\end{table}
Based on the analysis done in our previous papers \cite{Keus:2014jha,Keus:2015xya,Cordero-Cid:2016krd,Keus:2019szx}, we have chosen a number of BPs to represent different regions of parameter space in the model. As the aim of the paper is to test the model at colliders, we are focusing here on relatively light masses of DM particles, with $m_{S_1} \lesssim 80$ GeV. In this mass region, the I(2+1)HDM provides three distinctive types of benchmark scenarios, as follows.
\begin{enumerate}
\item \textbf{Scenario A}: with a large mass splittings, of order 50 GeV or so, between the DM candidate $S_1$ and all other inert particles, $m_{S_1} \ll m_{S_2}, m_{S_3}, m_{S_4}, m_{S^\pm_1}, m_{S^\pm_2}$. Scenarios of this type can be realised within the mass range $53$ GeV $ \leq m_{\rm DM} \leq 75 $ GeV in agreement with all theoretical and experimental constraints, provided the Higgs-DM coupling, $g_{h{\rm DM}}$, is relatively small.
\item \textbf{Scenario B}: with a small mass splitting, of order 20\% of $m_{\rm DM}$, between the DM and the next-to-lightest inert neutral particle, $m_{S_1} \sim m_{S_2} \ll m_{S_3},  m_{S_4}, m_{S^\pm_1}, m_{S^\pm_2}$. This choice also leads to a relatively small mass splitting between $S_3$ and $S_4$,  effectively separating the neutral sector into two groups, with each generation accompanied by a charged scalar. 
\item \textbf{Scenario C}: with all neutral particles close in mass, $m_{S_1} \sim m_{S_2} \sim m_{S_3} \sim  m_{S_4} \ll m_{S^\pm_1} \sim m_{S^\pm_2}$. Across the whole low and medium mass range, this scenario under-produces DM, due to the small mass splittings of the neutral inert particles which in turn strengthen the coannihilation channels, reducing the DM relic density.
\item \textbf{Scenario D}: which is essentially a scenario A with large $Z S_i S_j$ couplings of order 0.1, and therefore a smaller relic density.
\end{enumerate}

For each BP, we list the input parameters, i.e., masses of particles and all relevant couplings, following the convention:
\begin{eqnarray} 
&& \mathcal{L}_{\rm gauge} \supset g_{ZS_iS_j} Z_\mu (S_i \partial^\mu S_j - S_j \partial^\mu S_i), \label{gZSS}\\
&& \mathcal{L}_{\rm scalar} \supset \frac{v}{2}g_{S_i S_i h} h S_i^2+ v g_{S_i S_j h} h S_i S_j + v g_{S_i^\pm S_j^\mp h} h S_i^\pm S_j^\mp. \label{ghSS}
\end{eqnarray}
Table \ref{table-BPs} shows the input and derived parameters for each of the BPs.

\section{The effective $ZZZ$ vertex}
\label{ZZZ-section}

\subsection{The Lorentz structure and the $f_4^Z$ contribution}
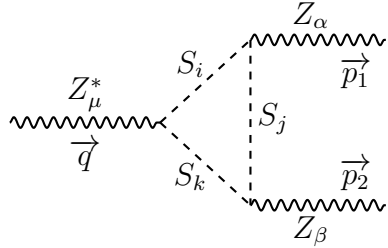
\begin{figure}[h!]
\begin{center}       
\begin{tikzpicture}[thick,scale=1.0]
\draw[photon] (0,0) -- node[black,above,xshift=0cm,yshift=0cm] {$Z^*_\mu$} (2,0);
\draw (0,0)  node[black,above,xshift=1cm,yshift=-0.8cm] {$\overrightarrow{q}$} (2,0);
\draw[dashed] (2,0) -- node[black,above,xshift=-0.2cm,yshift=-0.1cm] {$S_i$} (3.2,1.1);
\draw[dashed] (3.2,1.1) -- node[black,above,xshift=0.3cm,yshift=-0.3cm] {$S_j$} (3.2,-1.1);
\draw[dashed] (2,0) -- node[black,above,xshift=-0.2cm,yshift=-0.5cm] {$S_k$} (3.2,-1.1);
\draw[photon] (3.2,1.1) -- node[black,above,xshift=-0.1cm,yshift=0cm] {$Z_\alpha$} (5,1.1);
\draw (3.2,1.1)  node[black,above,xshift=1.4cm,yshift=-0.8cm] {$\overrightarrow{p_1}$} (5,1.1);
\draw[photon] (3.2,-1.1) -- node[black,above,xshift=-0.1cm,yshift=-0.7cm] {$Z_\beta$} (5,-1.1);
\draw  (3.2,-1.1)  node[black,above,xshift=1.4cm,yshift=0.05cm] {$\overrightarrow{p_2}$} (5,-1.1);
\end{tikzpicture}
\caption{The one-loop triangle diagram contributing to the $f_4^Z$ factor in the $ZZZ$ vertex, mediated by non-identical scalars $S_{i},S_{j},S_{k}$.}
\label{ZZZ-fig}
\end{center}
\end{figure}

The CP-violating weak basis invariants \cite{Lavoura:1994fv,Botella:1994cs,Gunion:2005ja,Grzadkowski:2014ada,Haber:2006ue,Haber:2015pua}, in particular the invariant which represents CP-violation in the mass matrix, contribute to the effective $ZZZ$ vertex.
This particular invariant is proportional to the mass splitting between the scalars which mediate the $ZZZ$ loop, shown in Figure \ref{ZZZ-fig}, the scalar-scalar-$Z$ couplings and inversely proportional to the scalar masses \cite{Lavoura:1994fv},
\be 
J_{CP} \, \propto \, \frac{|m^2_{S_i}-m^2_{S_j}|\,|m^2_{S_j}-m^2_{S_k}|\,|m^2_{S_k}-m^2_{S_i}|}{m^2_{S_i} m^2_{S_j} m^2_{S_k}} \,
| g_{ZS_iS_j}| \,| g_{ZS_jS_k}| \, |g_{ZS_kS_i}|,
\label{JCP}
\ee
where $i \neq j \neq k$, i.e., the scalars in the loop are non-identical.

In the context of the 2HDM, the CP-violating form factors for triple gauge boson couplings are known \cite{He:1992qh,Chang:1994cs,Chang:1993vv} and have been studied phenomenologically \cite{Hagiwara:1986vm,Gounaris:1999kf,Gounaris:2000dn,Baur:2000ae,Grzadkowski:2016lpv}.
Following the convention of \cite{Hagiwara:1986vm,Grzadkowski:2016lpv}, the Lorentz structure of the $ZZZ$ vertex
when the incoming $Z^*$ boson, characterised by momenta and Lorentz index ($q,\mu$), is assumed to be off-shell and the outgoing $Z$ bosons, characterised by ($p_1,\alpha$) and ($p_2,\beta$), are assumed to be on-shell, as shown in Figure~\ref{ZZZ-fig}, is reduced to 
\be
e\, \Gamma^{\alpha\beta\mu}_{ZZZ} = i\, e\,\frac{q^2 - m_Z^2}{m_Z^2}\, \biggl[ f_4^Z(q^\alpha g^{\mu\beta} + q^\beta g^{\mu\alpha}) + f_5^Z\epsilon^{\mu\alpha\beta\rho}(p_1 - p_2)_\rho \biggr],
\label{Gammaf4}
\ee
where $e$ is the proton charge. Also, it is assumed that $Z^*$ couples to a pair of light fermions $f \bar f$, hence, the terms proportional to the fermion mass have been neglected.
The dimensionless form factor $f_4^Z$ violates CP while $f_5^Z$ conserves CP. 
In our set-up, the $f_5^Z$ contributions are purely from the SM, while the scalar CP-violation contributes to $f_4^Z$ solely through the triangle diagram shown in Figure~\ref{ZZZ-fig} with $S_i S_j S_k$ in the loop, since the odd $Z_2$ charge of the inert sector forbids any other diagrams\footnote{For example, triangle diagrams where one neutral inert scalar is replaced by a neutral Goldstone $G_0$ or a $Z$ boson.}.

\begin{figure}[h!]
\centering
\includegraphics[width=7.5cm]{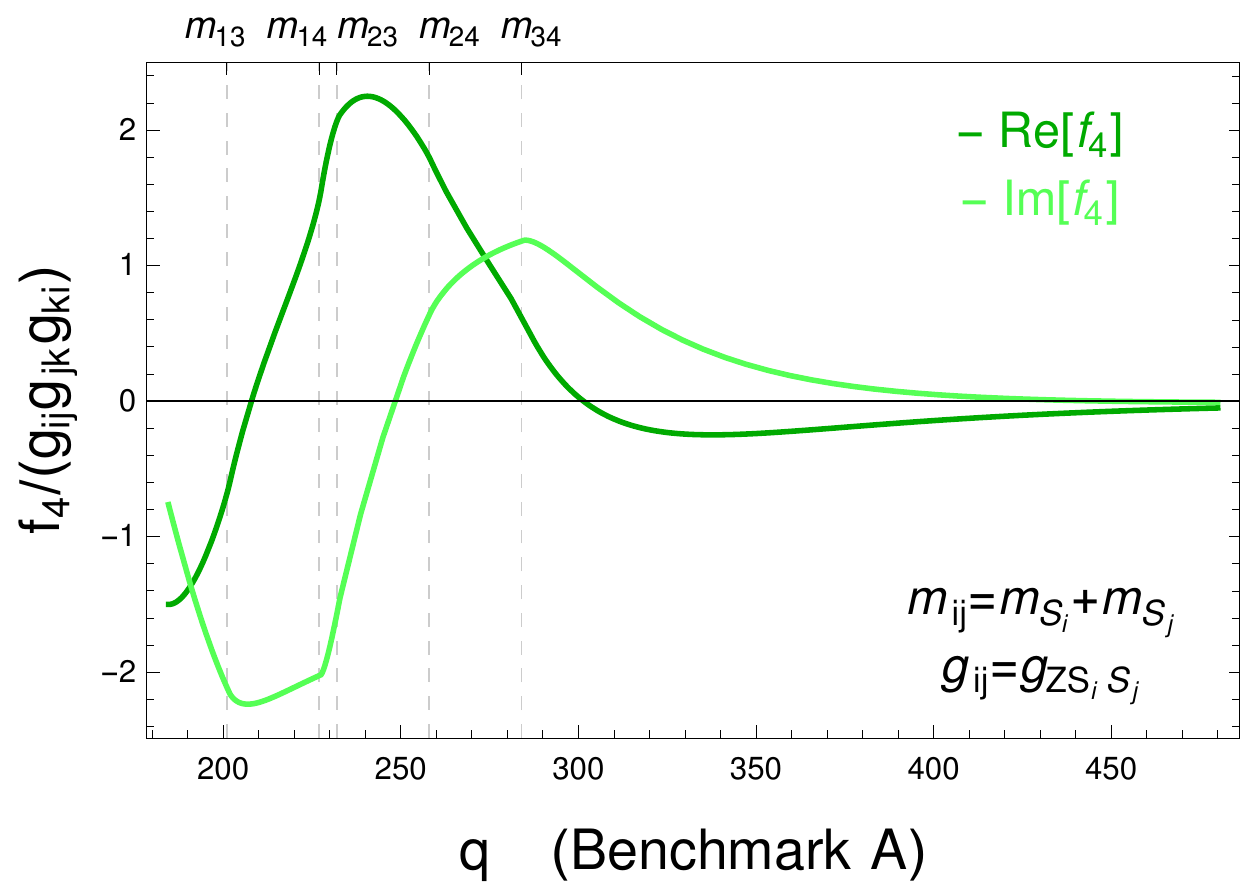}~
\includegraphics[width=7.5cm]{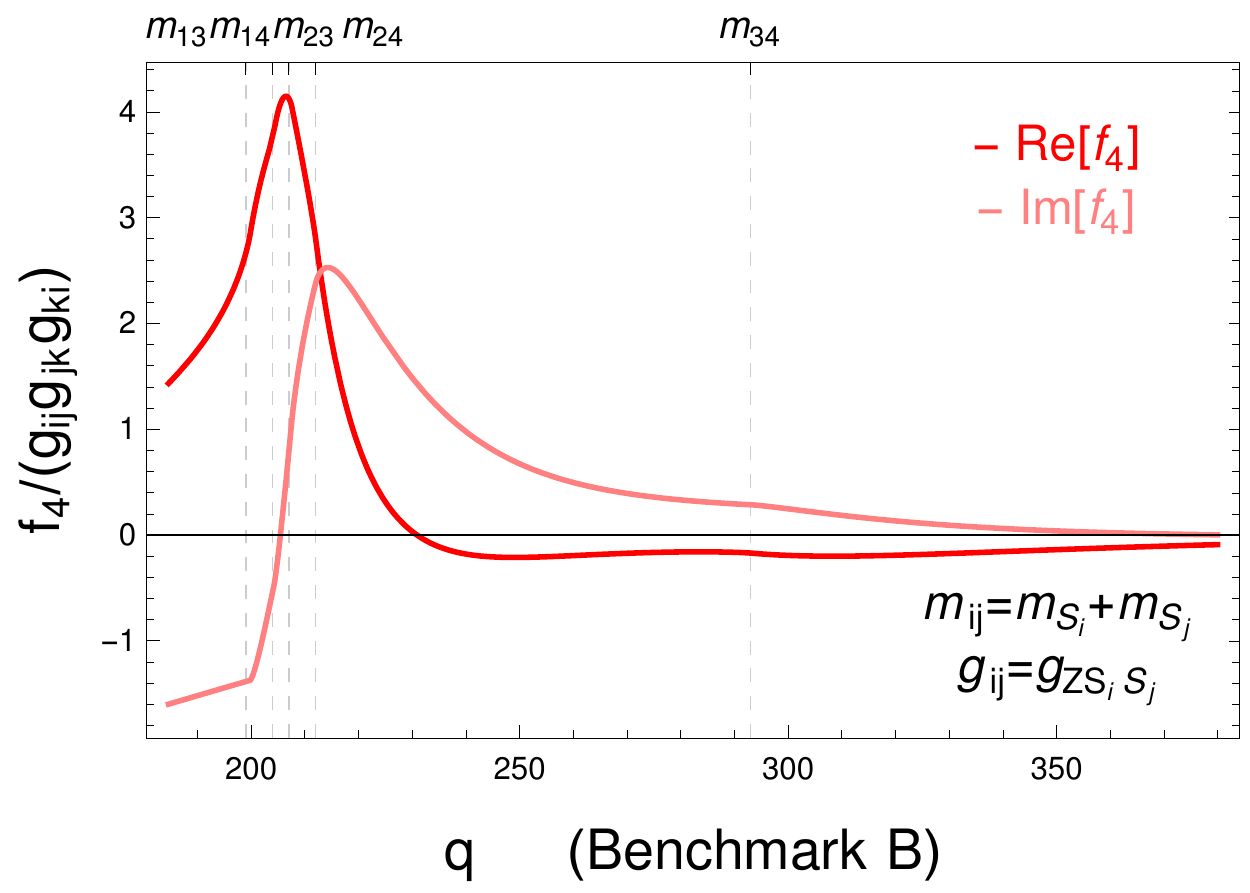}\\[1mm]
\includegraphics[width=7.5cm]{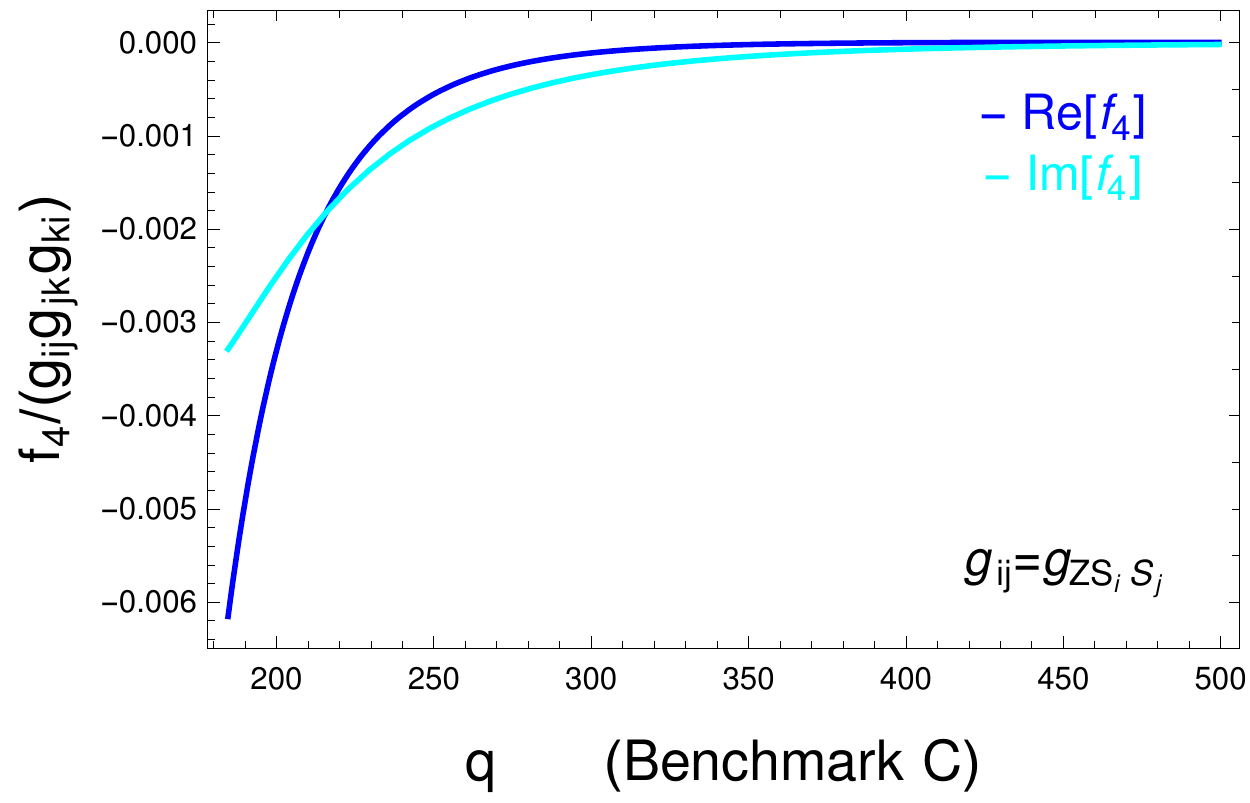}~
\includegraphics[width=7.5cm]{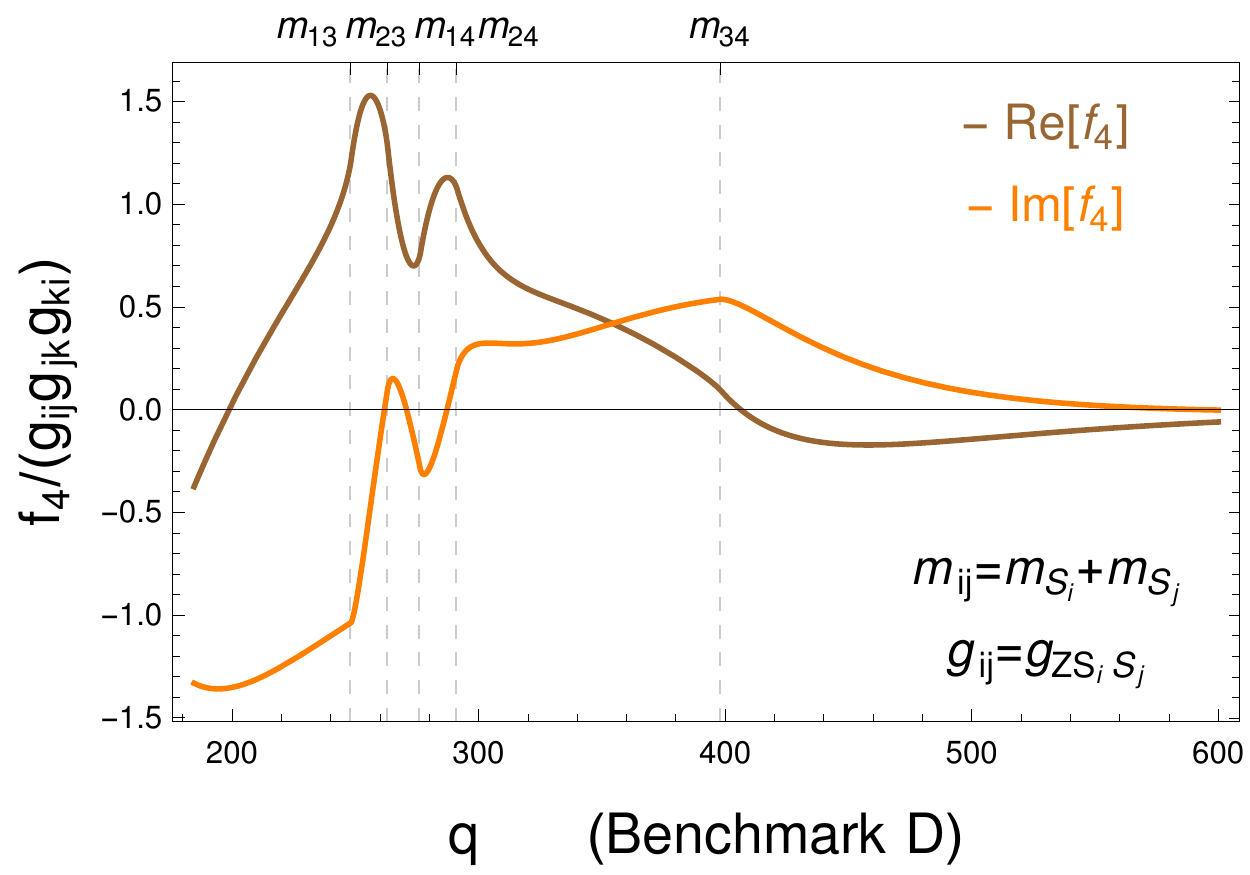}   
\caption{The $f_4^Z$ value (rescaled by the product of the three $ZS_iS_j$ couplings) in each BP, with respect to the momentum of the off-shell incoming $Z^*$ boson.}
\label{f4A}
\end{figure}

Using the package \texttt{LoopTools} \cite{Hahn:2010zi}, we calculate the total one-loop contribution to the $f_4^Z$ factor in our model to be given by a linear combination of the three-point tensor coefficient functions $C_{002}$ (in the \texttt{LoopTools} notation) as:
\be
f_4^Z =
\frac{m_Z^2}{2\pi^2\, e \, (q^2 - m_Z^2)} \,
| g_{ZS_2S_3}| \, | g_{ZS_1S_3}| \, |g_{ZS_1S_2}|  
\sum_{i,j,k}^4\epsilon_{ijk}C_{002}(m_Z^2,m_Z^2,q^2,m^2_i,m^2_j,m^2_k),
\label{f4}
\ee
where $m_{i,j,k}$ stands for the mass of the $S_{i,j,k}$ scalar.
Figure \ref{f4A} shows the value of $f_4^Z$ (rescaled by the product of the three $ZS_iS_j$ couplings) with respect to the momentum of the off-shell incoming $Z^*$ boson, $q$, for all our BPs. Here, for cases A, B and D, we have highlighted the mass thresholds inside the loop at $q=m_{ij}=m_{S_i}+m_{S_j}$. 
The mass thresholds in point C appear around 100 GeV which is well below the energy required for a $ZZ$ final state.
As expected from Eq. (\ref{JCP}), BPs with larger scalar mass splittings have a larger $f_4^Z$ contribution, namely points A, B and D, while small mass splittings lead to a small $f_4^Z$ contribution, as in point C.

\subsection{The $f \bar f \to Z^* \to Z Z$ cross section}

The expression in Eq. (\ref{Gammaf4}) can be extracted from the following effective Lagrangian describing the $V^*ZZ$ coupling ($V=\gamma,Z$) \cite{Gounaris:2000tb,Moyotl:2015bia,Azevedo:2018fmj}:
\be 
\mathcal{L}_{ZZZ^*} = -\frac{e}{m^2_Z}f_4^Z(\partial_\mu Z^{\mu\beta})Z_\alpha (\partial^\alpha Z_\beta),
\ee
where $Z_{\mu\nu} = \partial_\mu V_\nu - \partial_\nu V_\mu$. 

\begin{figure}
\centering
\includegraphics[scale=0.55]{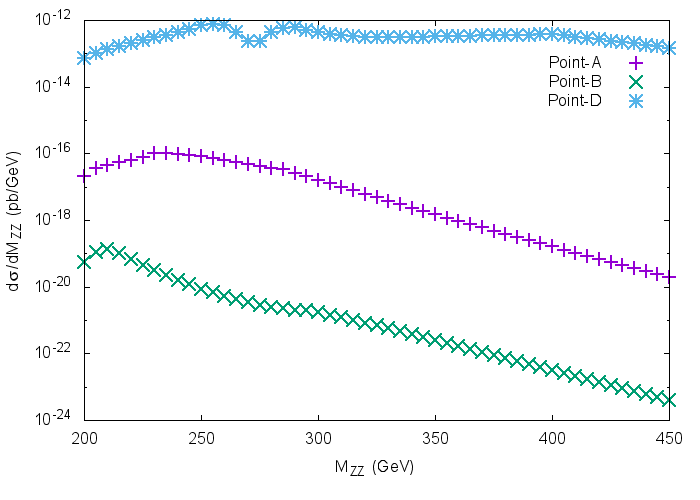}
\caption{The differential cross section  $d\sigma/dM_{ZZ}$ versus $M_{ZZ}$
for the $q\bar q \to Z^* \to Z Z$ process  for BPs A, B and D at the 14 TeV  LHC.}
 \label{sigmadat}
\end{figure}

Figure \ref{sigmadat} shows the differential cross section at the LHC for the $q\bar q \to Z^* \to Z Z$ process,   i.e.,  $d\sigma/dM_{ZZ}$ versus $M_{ZZ}$, obtained with \texttt{CalcHEP} \cite{Belyaev:2012qa} for BPs A, B and D. We do not show the cross section plots for BP C, since the corresponding $f_4^Z$  is very small. Here, we have used  $\sqrt{s}=14$ TeV as collider energy and the CTEQ6L1 Parton Distribution Functions (PDFs) \cite{Stump:2003yu} with renormalisation/factorisation scale set equal to $M_{ZZ}$.
Comparing Figures \ref{f4A} and \ref{sigmadat}, it is evident that the cross section plots represent the pattern of the $f_4^Z$ ones for each benchmark scenario with $|f_4^Z| = \sqrt{{\Re f_4^Z}^2+ {\Im f_4^Z}^2}$. 

Note that the $q\bar{q} \to ZZ$ process has a large tree-level contribution from the SM whose interference with the one-loop $ZZZ$ process might be observable. However, this interference term is noted to be zero in \cite{Gounaris:1999kf}. We have verified this result by iteratively applying the Dirac equation on the interference term.

\begin{figure}
\centering
\includegraphics[scale=0.55]{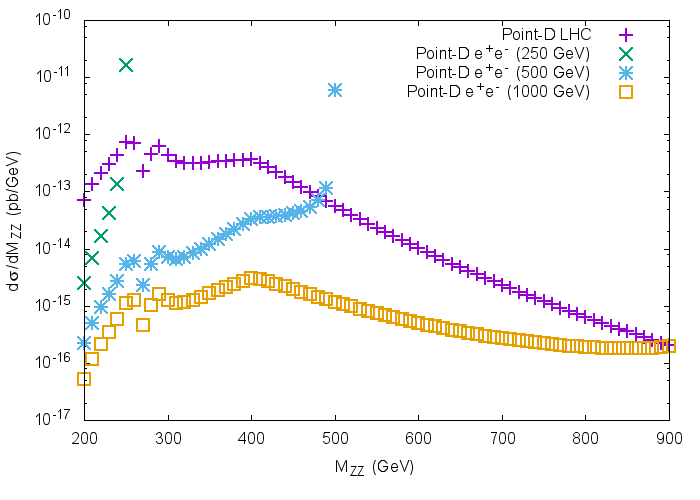}
\caption{The differential cross section  $d\sigma/dM_{ZZ}$ versus $M_{ZZ}$
for the $f \bar f \to Z^* \to Z Z$ process  for BP D at the  14 TeV LHC ($f=q$) and a lepton collider ($f=e$)  with different energies.}
\label{ee-collider}
\end{figure}

Figure \ref{ee-collider} compares the obtained $f \bar f \to Z^* \to Z Z$ cross section at the LHC (where $f=q$) and at a lepton collider (where $f=e$). While the result for the hadron collider was obtained considering an energy of 14 TeV, for the lepton collider we considered the energies of 250, 500 and 1000 GeV, which are the values proposed for future $e^+ e^-$ colliders such as the Future Circular Collider in $e^+e^-$ mode (FCC-ee), International Linear Collider (ILC), Compact Linear Collider (CLiC) or Circular Electron-Positron Collider (CEPC), see \cite{Craig:2017gzf} for a comparison of their physics potential. The selected electron/positron PDFs in \texttt{CalcHEP} are the default ones (and we do not include bremsstrahlung effects). Herein, it is remarkable to notice that the LHC distribution generally has a much larger cross section than those at leptonic colliders, except for $M_{ZZ}\approx \sqrt s_{e^+e^-}$ (which is natural, as without electron/positron PDFs the distribution would be a $\delta$-function at the lepton collider energy\footnote{A similar effect does not occur at the LHC, where the incoming (anti)quark pair is confined inside the proton beams.}). However, very large luminosities would be required to observe any event at any of these colliders. This is nonetheless a rather novel result, as previous literature exclusively concentrated on $e^+e^-$ colliders, thus overlooking the fact  that the LHC generally has more sensitivity to the CP-violating contributions entering the $ZZZ$ vertex. 
Finally, here, we have illustrated this phenomenology for the case of BP D which has the largest cross section amongst the studied BPs due to its large $g_{ZS_iS_j}$ couplings, but the same pattern is also seen for the other cases.

\section{CP-violating asymmetries}
\label{asymmetry-section}

In an $f \bar f \to Z Z$ process, the helicities/polarisations of the $ZZ$ pair can be measured statistically from the angular distributions of their decay products.
If the helicities/polarisations of the $Z$ bosons are known, one could define CP-violating observables for the $Z Z$ state to test CP-violation at future colliders \cite{Chang:1994cs,Chang:1993vv,Grzadkowski:2016lpv,Djouadi:2007ik,Lebrun:2012hj,Gounaris:1991ce}.

These CP-violating observables are defined as differential asymmetries, assuming that both the momenta and helicities of the $ZZ$ pair can be determined (as explained).
Since our goal is to measure the CP-violating form factor $f_4^Z$, these asymmetries will (to leading order) be proportional to $f_4^Z$.

One can express the cross section $\sigma$ of the $f \bar f \to Z Z$ process as
\be 
\sigma(f_\delta \bar{f}_{\bar{\delta}} \to Z_{\eta} Z_{\bar\eta})
~\equiv ~
\sigma_{\eta,\bar\eta}=\sum_{\delta,\bar\delta} 
{\cal M}^{\delta,\bar\delta}_{\eta,\bar\eta}\,[\Theta]\,
{{\cal M}^\star}^{\delta,\bar\delta}_{\eta,\bar\eta}\,[\Theta],
\label{sigma-defined}
\ee
where $\delta, \bar\delta$ are the helicities of the incoming $f, \bar f$ and $\eta, \bar\eta$ are the helicities of the outgoing $Z Z$ pair, respectively \cite{Chang:1994cs}. 
Following from Eq. (\ref{Gammaf4}), the helicity amplitude $\mathcal{M}$ is given as
\be 
\mathcal{M}_{f \bar f \to Z Z} = \frac{1}{q^2-m_Z^2} \, \Gamma^{\mu \alpha \beta}_{ZZZ} \, \epsilon^\alpha(p_1 )\, \epsilon^\beta(p_2 ) j^\mu (q),
\ee
where $\epsilon^\alpha(p_1 )$ and $\epsilon^\beta(p_2 )$ are the polarisation vectors of the two outgoing on-shell $Z$ bosons with four momenta $p_1$ and $p_2$, respectively. The momentum of the off-shell $Z^*$ boson is characterised by $q=p_1+p_2$ and the fermionic current with which it connects to the Lagrangian is denoted by $j^\mu$. 
In the limit where the fermions are assumed to be massless, the $j^\mu$ current is conserved, $q_\mu \, j^\mu =0$.

In a lepton collider, the angle $\Theta$ is defined as the angle 
between, e.g.,  the incoming $e^{-}$ beam direction and the $Z$ whose helicity is given by the first index $\eta$. 
In a hadron collider, we make use of the event boost in the laboratory frame to determine the direction of the incoming particle, i.e., as the boost direction identifies with that of the incoming quark, with respect to which the angle $\Theta$ is then  measured. Hence, the forthcoming asymmetries, normally studied at lepton colliders, can also be exploited at the LHC.

Here, we introduce three observable asymmetries, namely $A^{ZZ}, \widetilde{A}^{ZZ}$ and ${A''}^{ZZ}$. Since the two $Z$ bosons in the final state are indistinguishable, for the observation of these asymmetries, one studies the forward hemisphere where one defines the $A_1$ asymmetry. Then, by studying the backward hemisphere, one defines the $A_2$ asymmetry. If the asymmetries in the two hemispheres are not equal, i.e. $A_1-A_2 \neq 0$, one can confidently claim that the model is CP-violating.

\subsection{Asymmetries $A_1^{ZZ}$ and $A_2^{ZZ}$}

The $A_1^{ZZ}$ and $A_2^{ZZ}$ asymmetries are defined as 
\be 
A_1^{ZZ} \equiv \frac{\sigma_{+,0}-\sigma_{0,-}}{\sigma_{+,0}+\sigma_{0,-}}, \qquad
A_2^{ZZ} \equiv \frac{\sigma_{0,+}-\sigma_{-,0}}{\sigma_{0,+}+\sigma_{-,0}},
\ee 
where $\sigma_{\eta,\bar{\eta}}$, as defined in Eq. (\ref{sigma-defined}), is the unpolarised beam cross section for the production of $ZZ$ with helicities $\eta$ and $\bar{\eta}$. 
With this definition, $A_1^{ZZ}$ and $A_2^{ZZ}$ are calculated to be 
\bea  
&& A_1^{ZZ} =
-4 \beta \gamma^4 \left[(1+\beta ^2)^2-(2 \beta \cos \Theta)^2 \right]
{\cal F}_1(\beta,\Theta)\,\Im f_4^Z, 
\nonumber\\[2mm]
&& A_2^{ZZ} = A_1^{ZZ}\left(\cos \Theta\to-\cos \Theta\right), 
\label{A12zz}
\eea 
to the lowest order in $f_4^Z$, where $\gamma=\sqrt{s}/(2m_Z)$ and $\beta^2=1-\gamma^{-2}$.
The prefactor ${\cal F}_1(\beta,\Theta)$ is defined as
\be 
{\cal F}_1(\beta,\Theta)
=\frac{N_0+N_1\cos\Theta+N_2\cos^2\Theta+N_3\cos^3\Theta}
{D_0+D_1\cos \Theta+D_2\cos^2 \Theta+D_3\cos^3 \Theta+D_4\cos^4 \Theta},
\ee
with the following coefficients
\begin{subequations}
\begin{align}
\xi_1&=\sin\theta_W\cos\theta_W(1-6\sin^2\theta_W+12\sin^4\theta_W), &\,
\xi_2&=16\sin^7\theta_W\cos\theta_W,
\nonumber\\[1mm]
\xi_3&=1-8\sin^2\theta_W+24\sin^4\theta_W-32\sin^6\theta_W, &\,
\xi_4&=32\sin^8\theta_W,
\nonumber\\[3mm]
N_0&=\left(1+\beta ^2\right) \xi _1, &\,
N_1&=-2 \beta ^2 \left(\xi _1-\xi _2\right),
\nonumber\\[1mm]
N_2&=\left(\beta ^2-3\right) \xi _1, &\,
N_3&=2 \left(\xi _1-\xi _2\right),
\nonumber\\[3mm]
D_0&=\left(1+\beta ^2\right)^2 \left(\xi _3+\xi _4\right), &\,
D_1&=2 \left(1-\beta ^4\right) \xi _3,
\nonumber\\[1mm]
D_2&=-\left(3+6 \beta ^2-\beta ^4\right) \left(\xi _3+\xi _4\right), &\,
D_3&=-4 \left(1-\beta ^2\right) \xi _3,
\nonumber\\[1mm]
D_4&=4 \left(\xi _3+\xi _4\right).
\nonumber
\end{align}
\end{subequations}

\begin{figure}[h!]
\begin{center}
\includegraphics[scale=0.54]{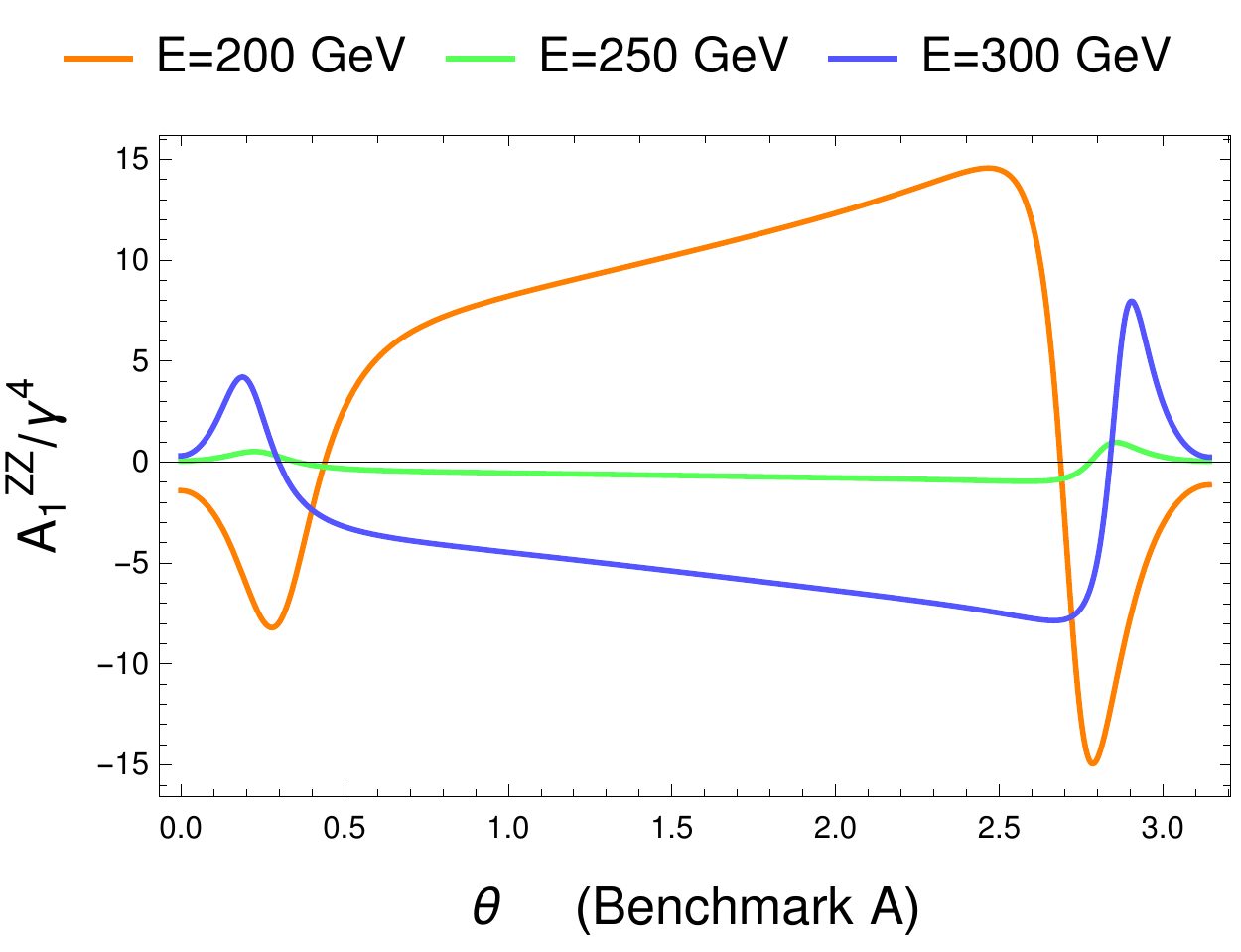}~~~
\includegraphics[scale=0.54]{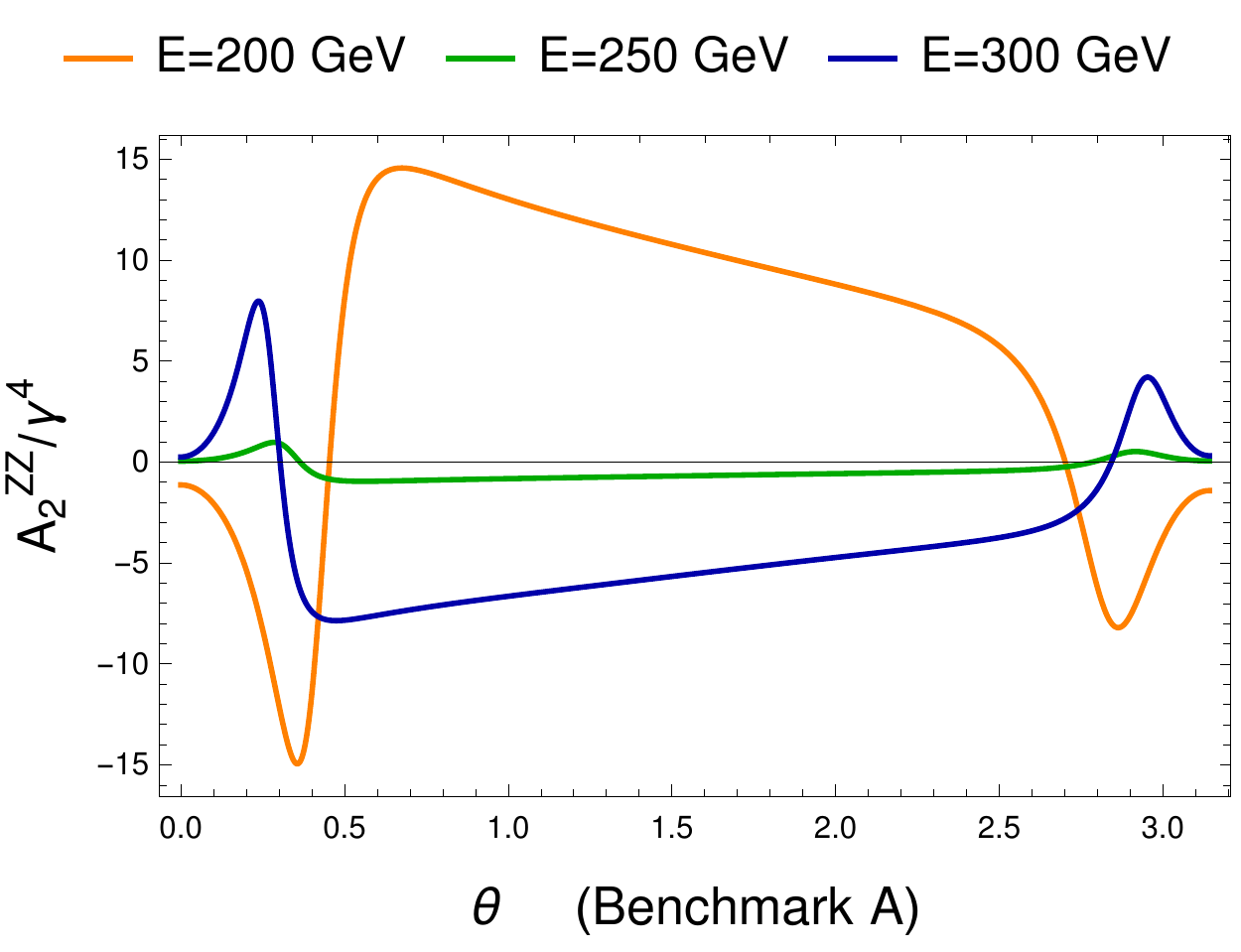}\\
\includegraphics[scale=0.54]{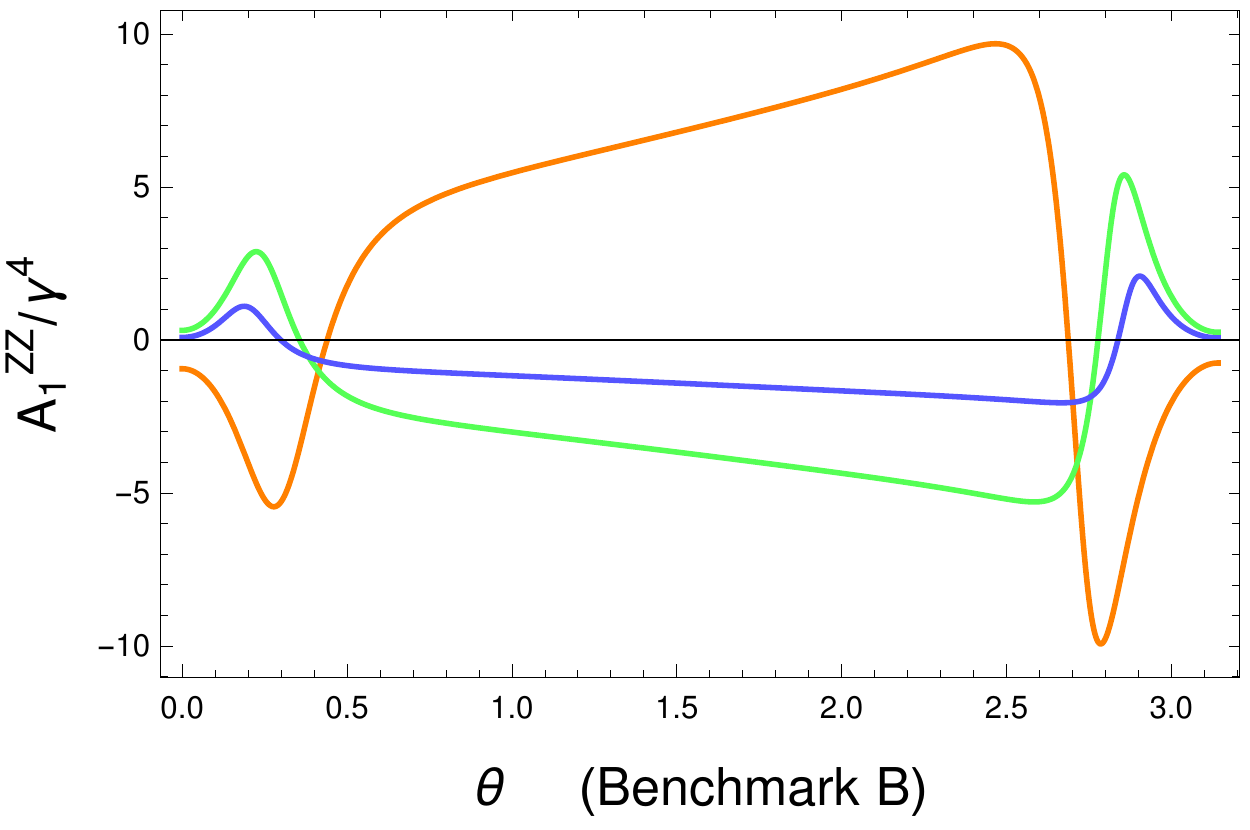}~~~
\includegraphics[scale=0.54]{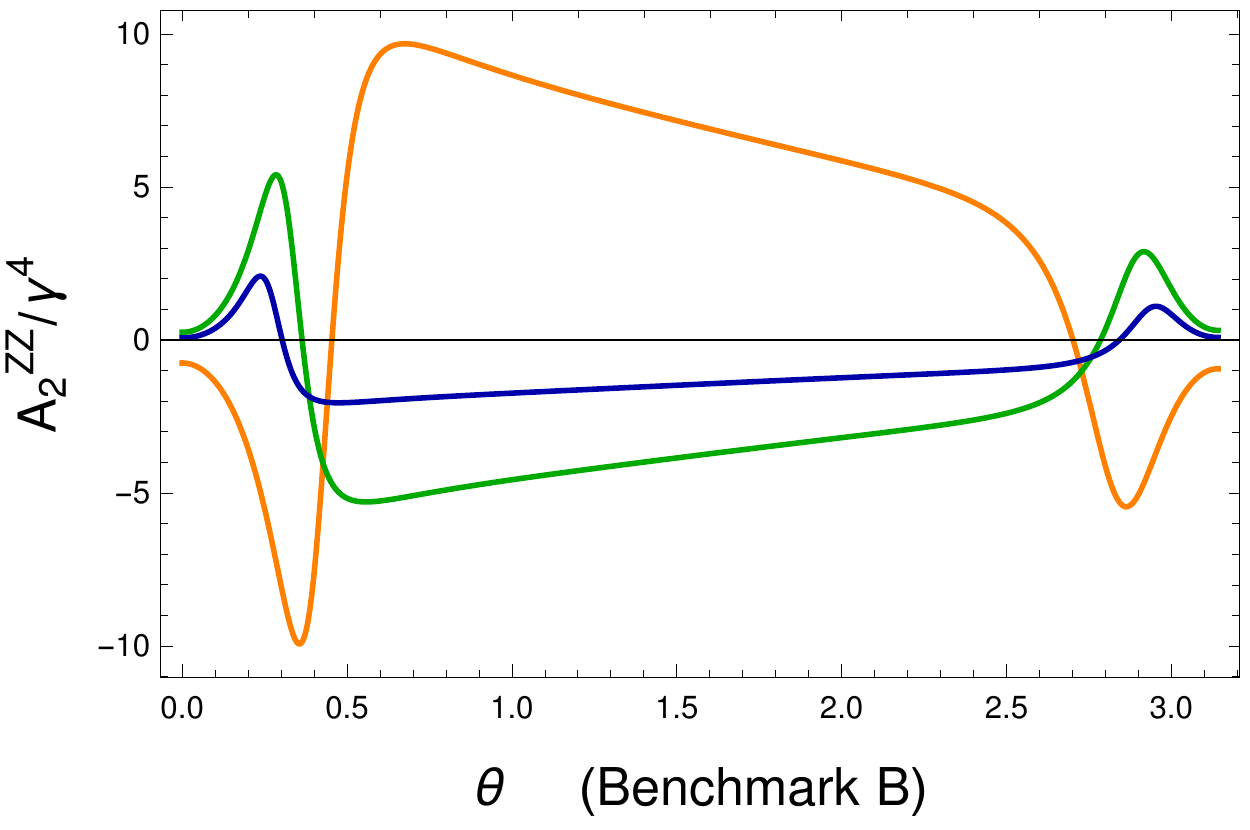}\\
\includegraphics[scale=0.54]{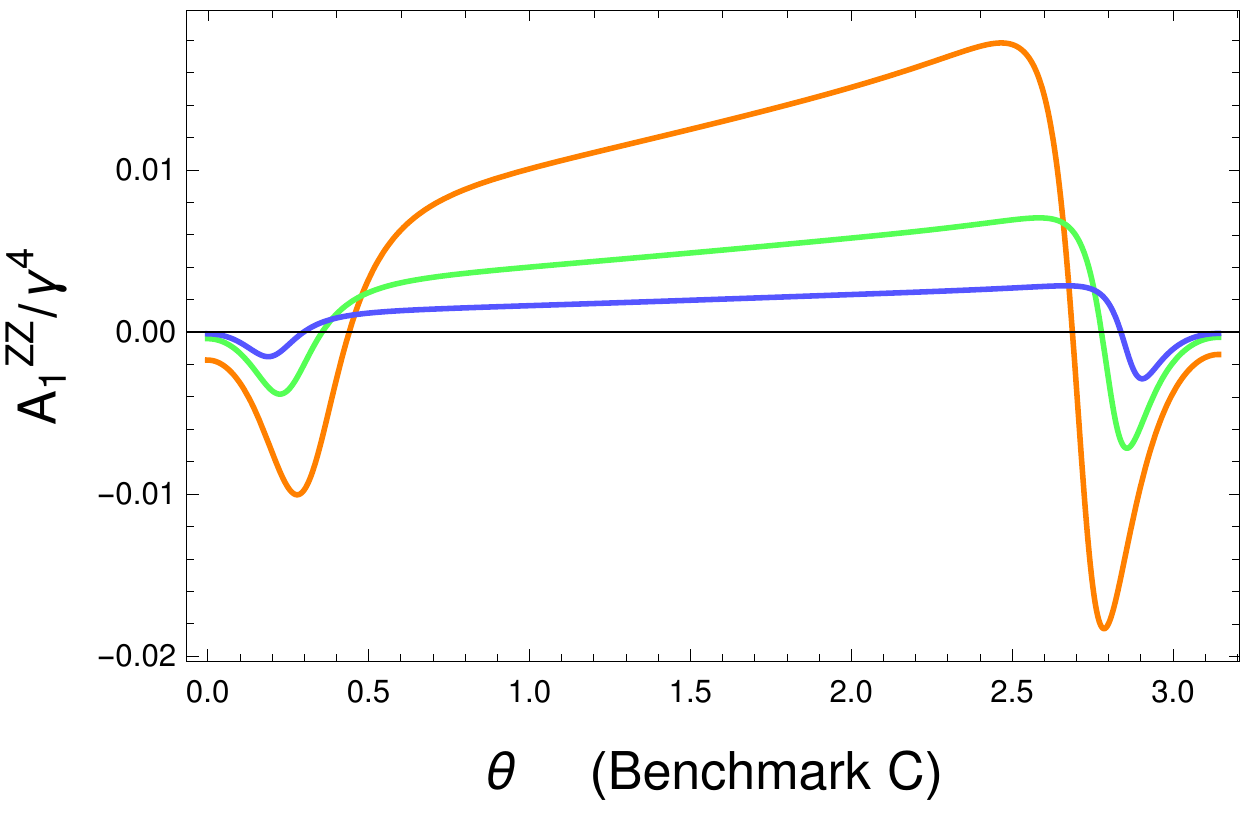}~~~
\includegraphics[scale=0.54]{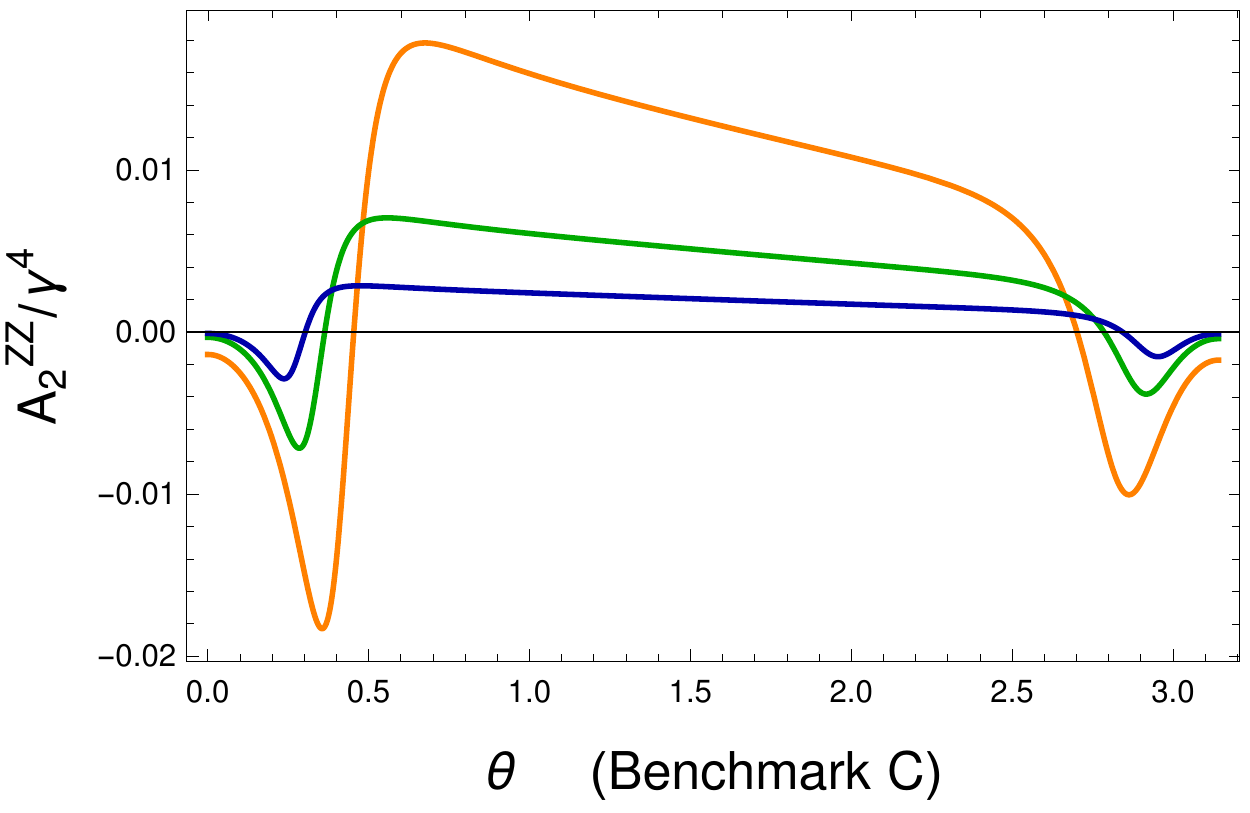}\\
\includegraphics[scale=0.54]{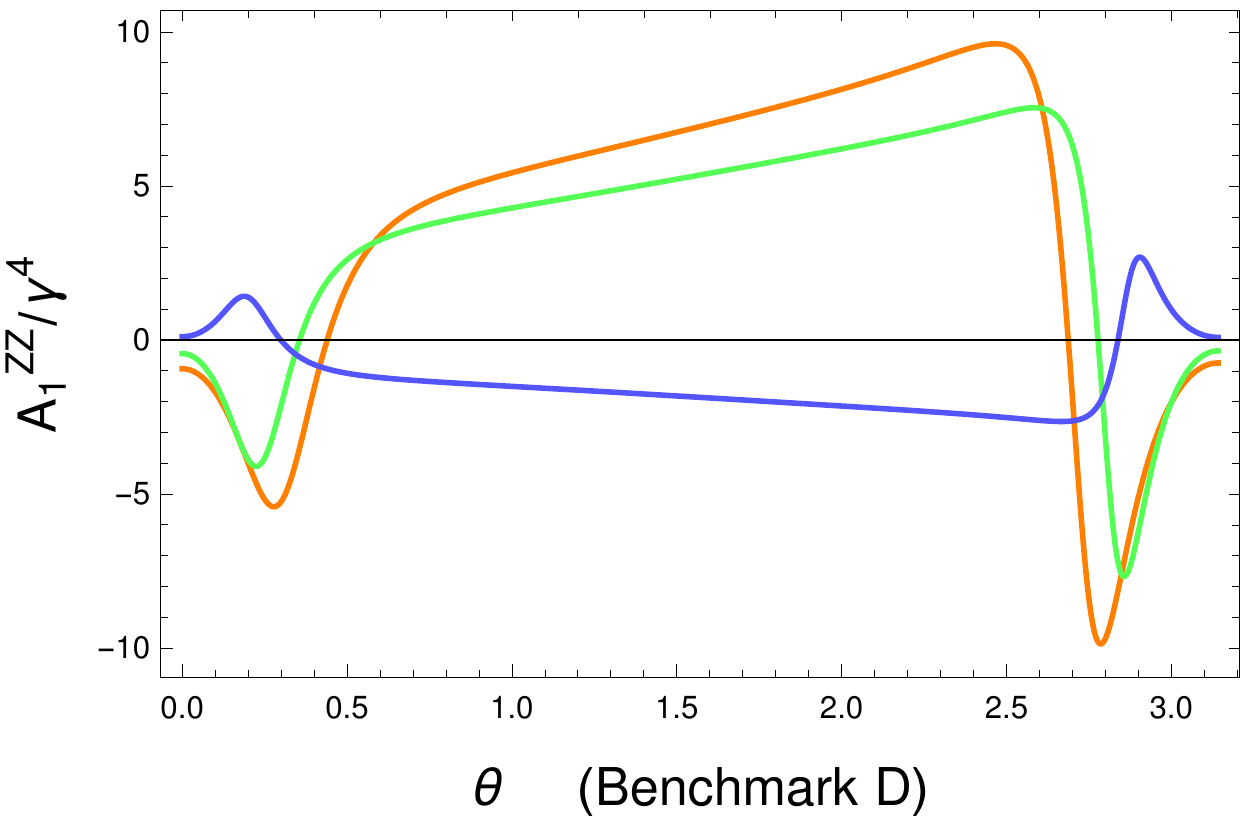}~~~
\includegraphics[scale=0.54]{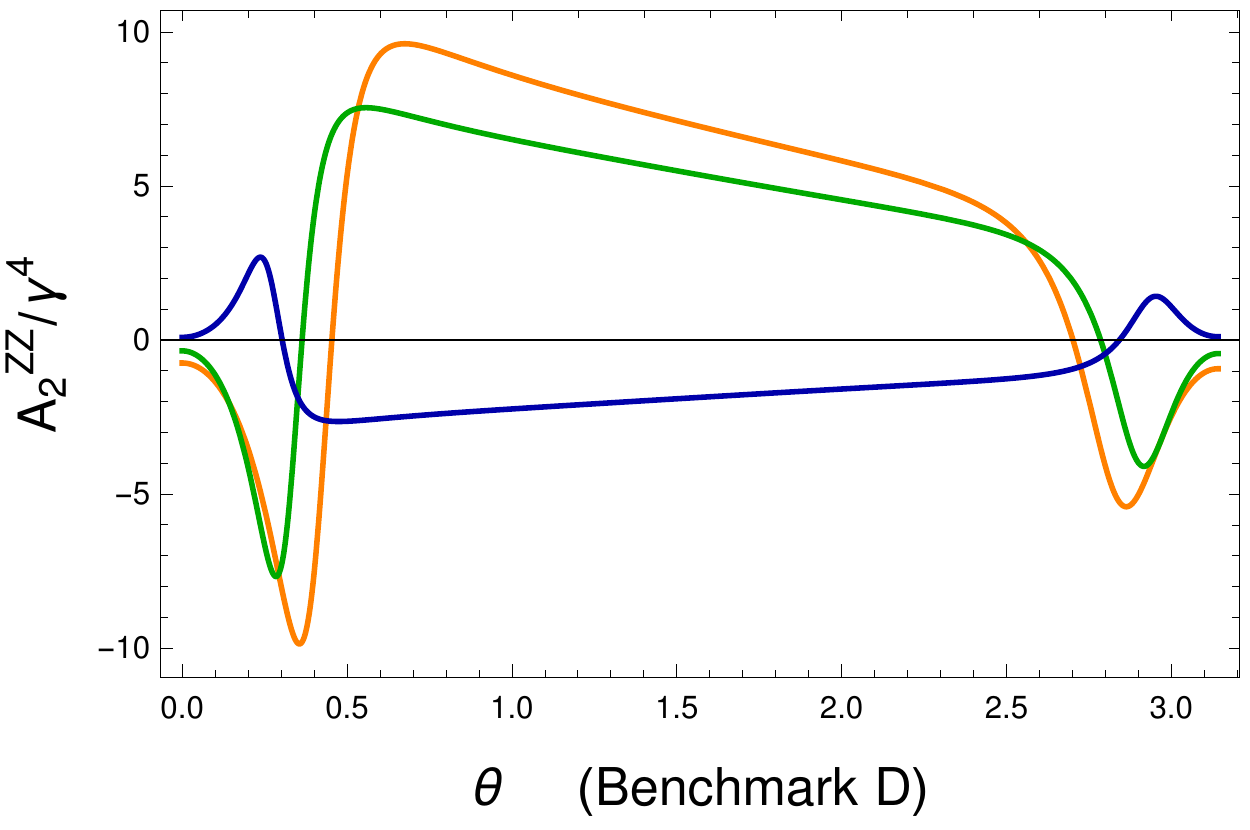}
\caption{The asymmetries $A_1^{ZZ}(\Theta)$ and $A_2^{ZZ}(\Theta)$  as functions of $\Theta$ for three beam energies $E$ as indicated (in GeV).}
\label{A1A2-fig}
\end{center}
\end{figure}

For all our BPs, we show these asymmetries in Figure~\ref{A1A2-fig}.

\subsection{Asymmetries $\widetilde{A}^{ZZ}_1$ and $\widetilde{A}^{ZZ}_2$}

Other CP-violating observables are the $\widetilde{A}^{ZZ}_1$ and $\widetilde{A}^{ZZ}_2$ asymmetries, defined as
\be 
\widetilde{A}^{ZZ}_1 \equiv \frac{\sigma_{+,0}+\sigma_{0,+}-\sigma_{0,-}-\sigma_{-,0}}{\sigma_{+,0}+\sigma_{0,+}+\sigma_{0,-}+\sigma_{-,0}} , 
\qquad
\widetilde{A}^{ZZ}_2 \equiv \frac{\sigma_{+,0}-\sigma_{0,+}-\sigma_{0,-}+\sigma_{-,0}}{\sigma_{+,0}+\sigma_{0,+}+\sigma_{0,-}+\sigma_{-,0}}.
\label{AAtilde}
\ee

Calculating these asymmetries to leading order in $f_4^Z$ reduces their expressions to
\bea
\widetilde{A}^{ZZ}_1
&=&
\biggl[
\frac{-2\beta\gamma^4[(1+\beta^2)^2-(2\beta\cos\Theta)^2][1+\beta^2-(3-\beta^2)\cos^2\Theta]}
{(1+\beta^2)^2-(3+6\beta^2-\beta^4)\cos^2\Theta+4\cos^4\Theta}
\biggr] \, \xi \, \Im f_4^Z ,
\nonumber\\[2mm]
\widetilde{A}^{ZZ}_2
&=&
\biggl[ \frac{-2\beta  \gamma ^4  \cos \Theta[(1+\beta^2)^2-(2\beta\cos\Theta)^2] \left(\beta ^2- \cos ^2\Theta \right) }
{\left(1+\beta ^2\right)^2-\left(3+6 \beta ^2-\beta ^4\right) \cos ^2\Theta +4 \cos ^4\Theta }
\biggr] \,\widetilde{\xi} \, \Im f_4^Z ,
\label{AAtilde-reduced} 
\eea
where we have defined $\xi$ and $\widetilde{\xi}$ to be
\bea
\xi&=&\frac{2\sin\theta_W\cos\theta_W(1-6\sin^2\theta_W+12\sin^4\theta_W)}{1-8\sin^2\theta_W+24\sin^4\theta_W-32\sin^6\theta_W+32\sin^8\theta_W},
%\simeq 1.65,
\nonumber\\[2mm]
\widetilde{\xi}&=&\frac{-4 \sin \theta _W \cos \theta _W\left(1-6 \sin ^2\theta _W+12 \sin ^4\theta _W-16 \sin ^6\theta _W\right) }
{1-8 \sin ^2\theta _W+24 \sin ^4\theta _W-32 \sin ^6\theta _W+32 \sin ^8\theta _W}
%\simeq -0.78 .
\label{xis}.
\eea

\begin{figure}[!t]
\begin{center}
\includegraphics[scale=0.54]{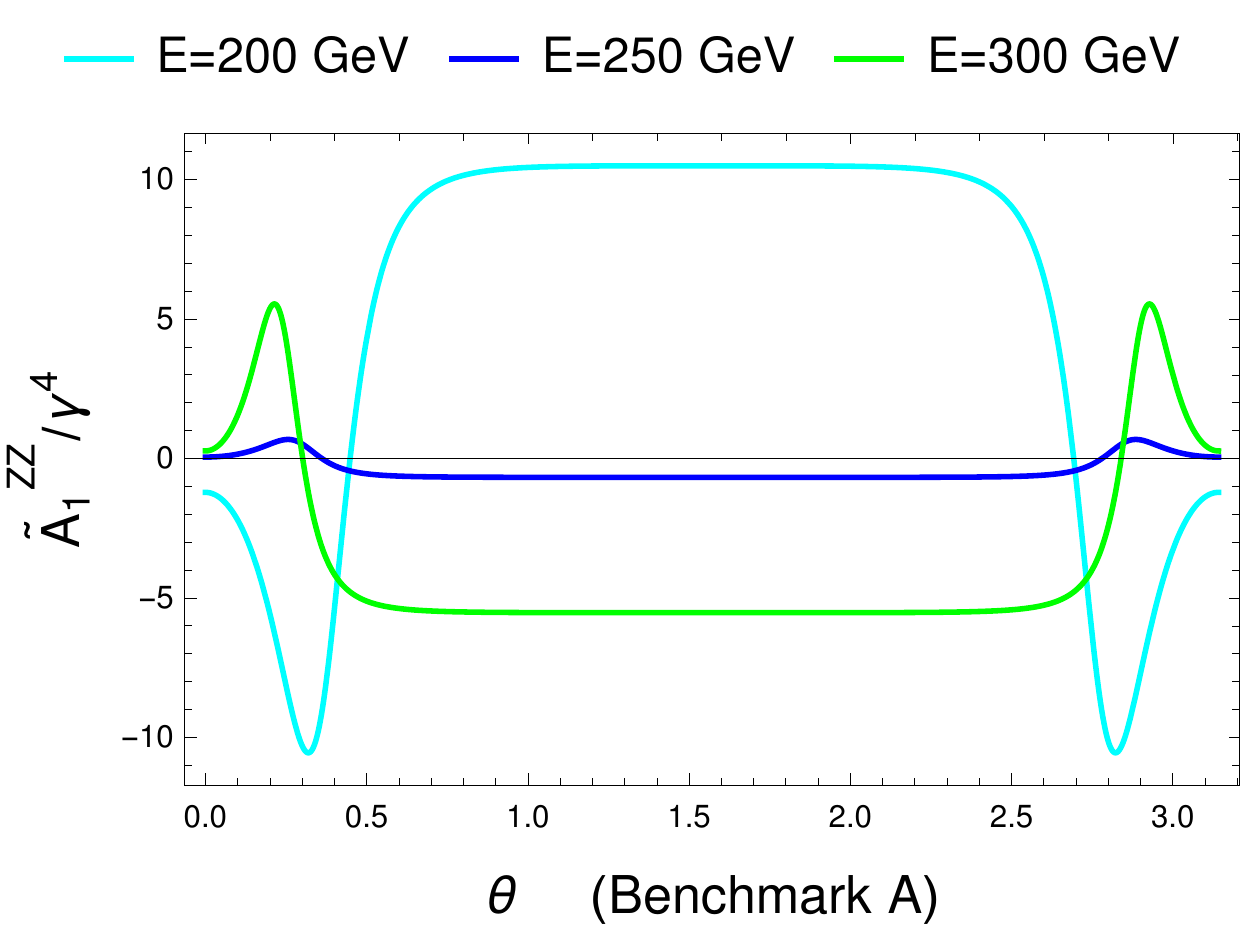}~~~
\includegraphics[scale=0.54]{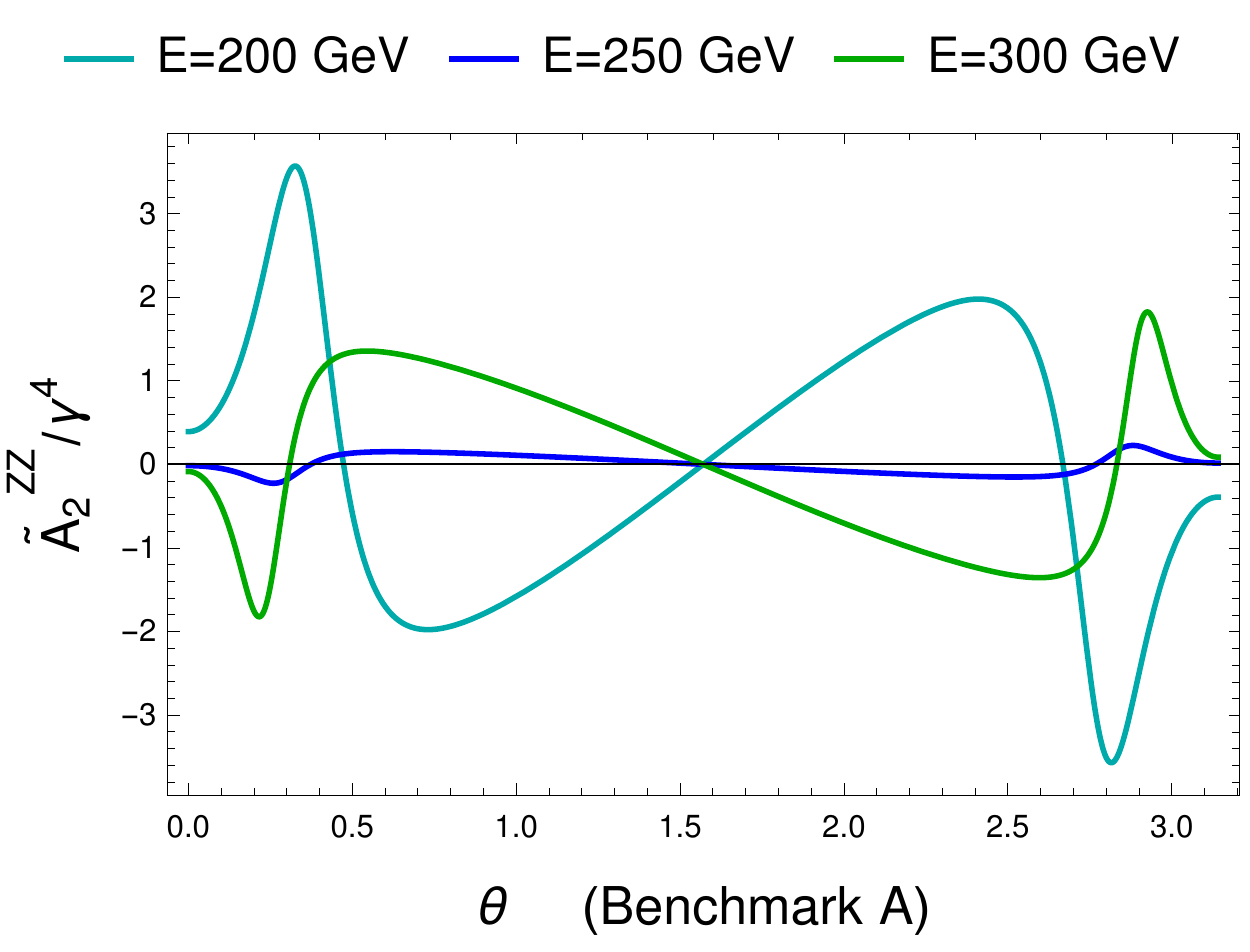}\\
\includegraphics[scale=0.54]{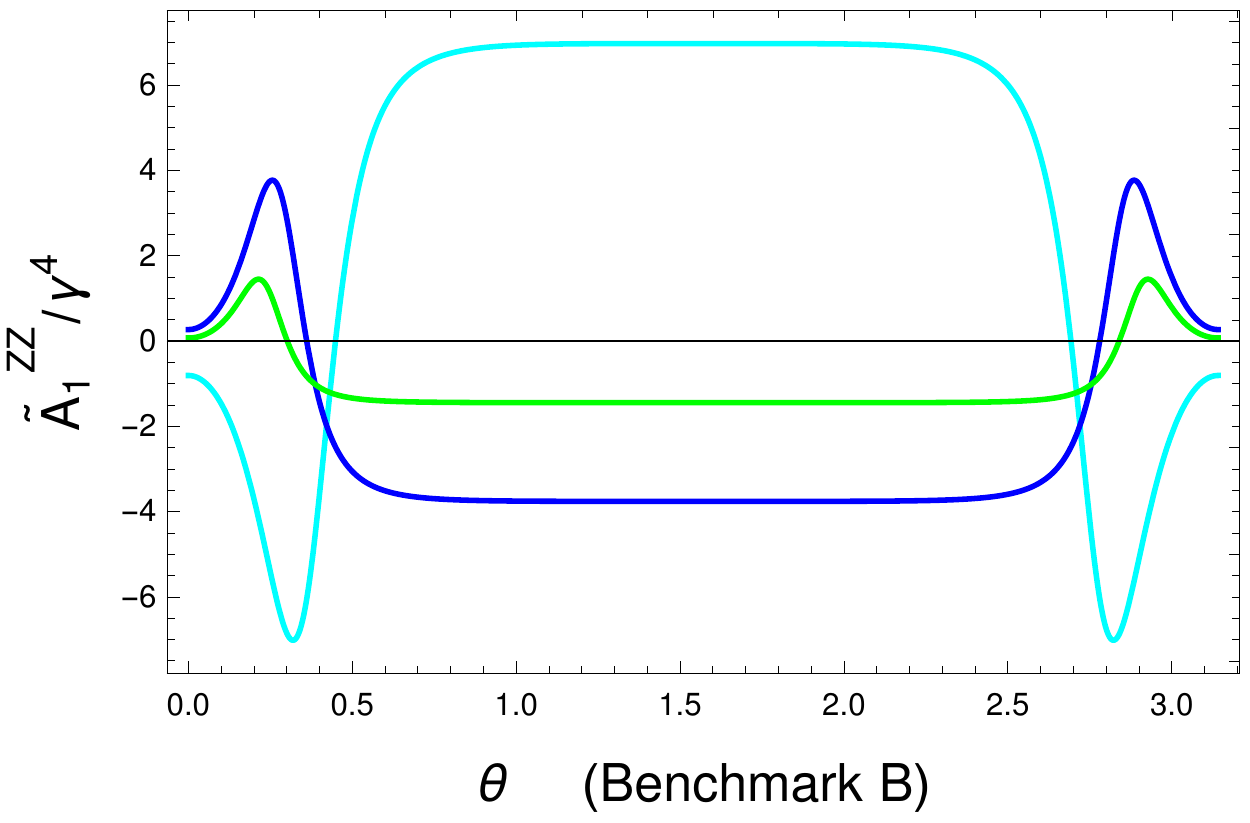}~~~
\includegraphics[scale=0.54]{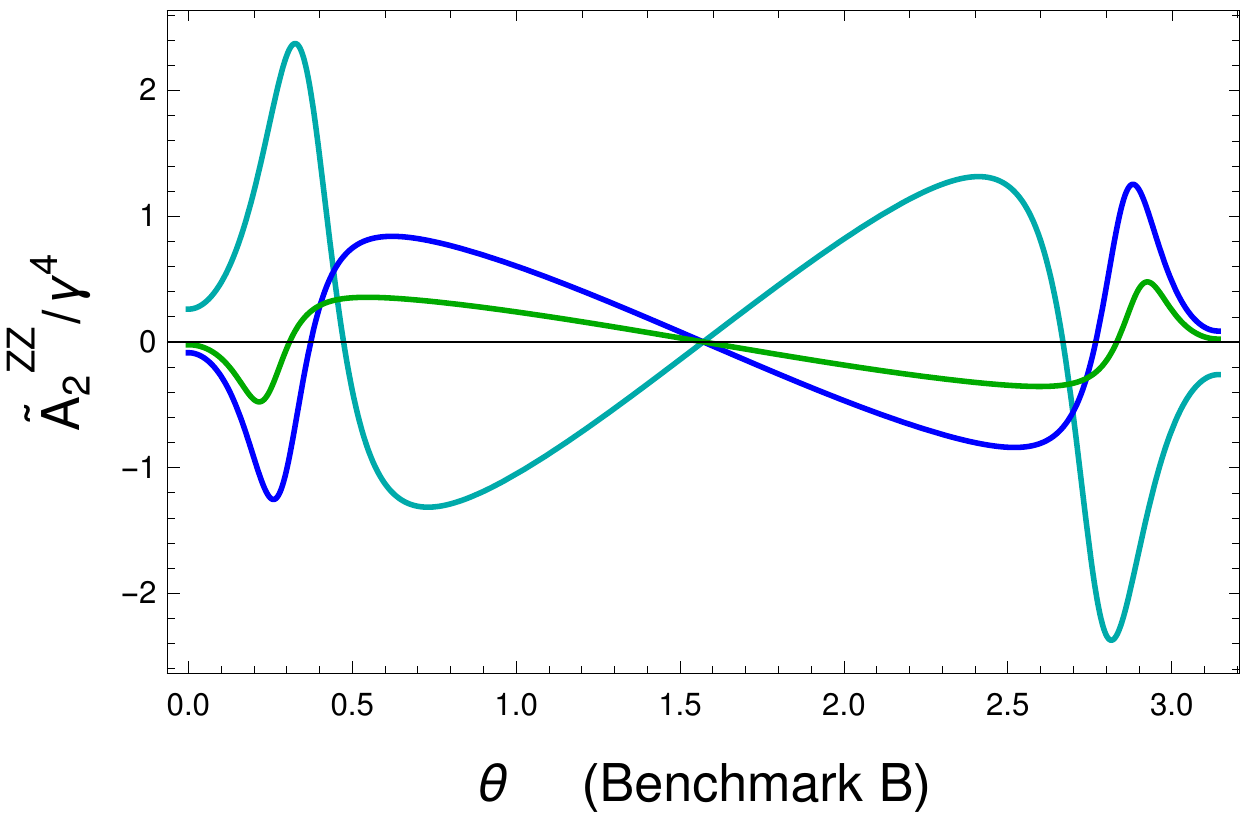}\\
\includegraphics[scale=0.54]{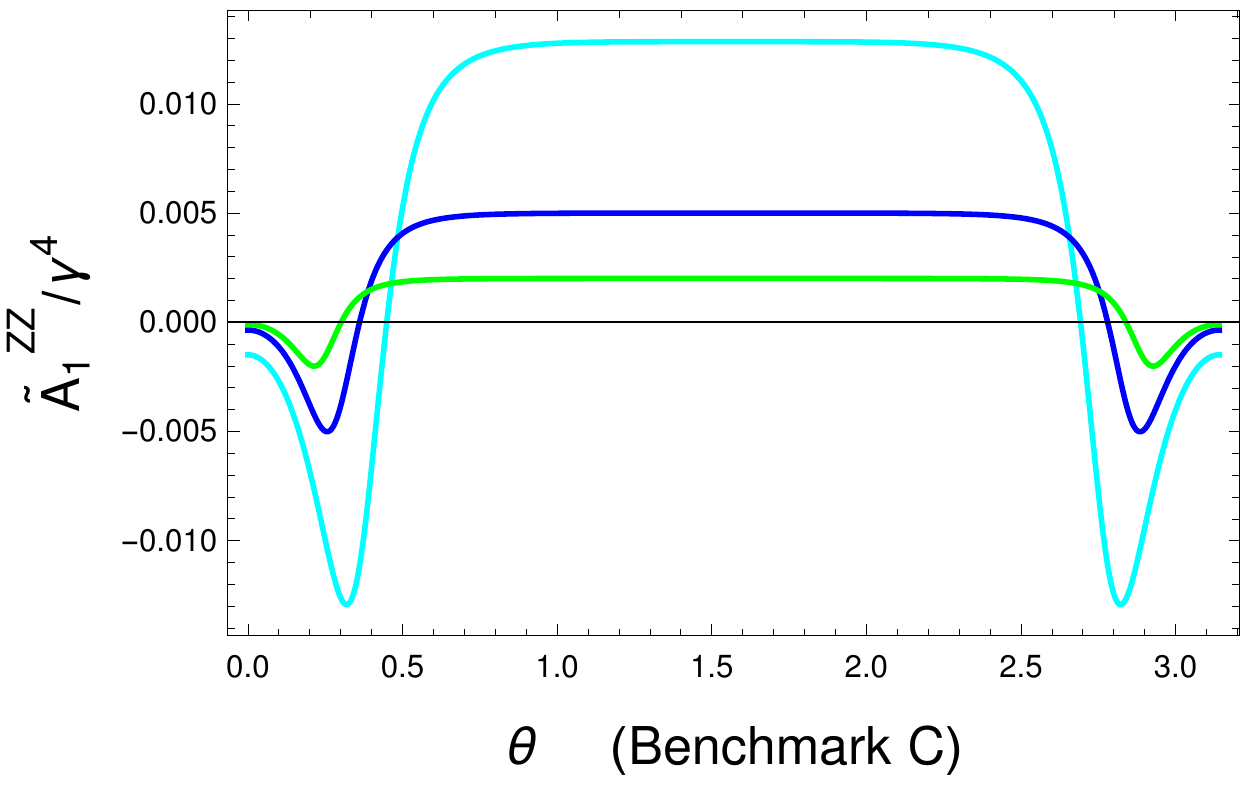}~~~
\includegraphics[scale=0.54]{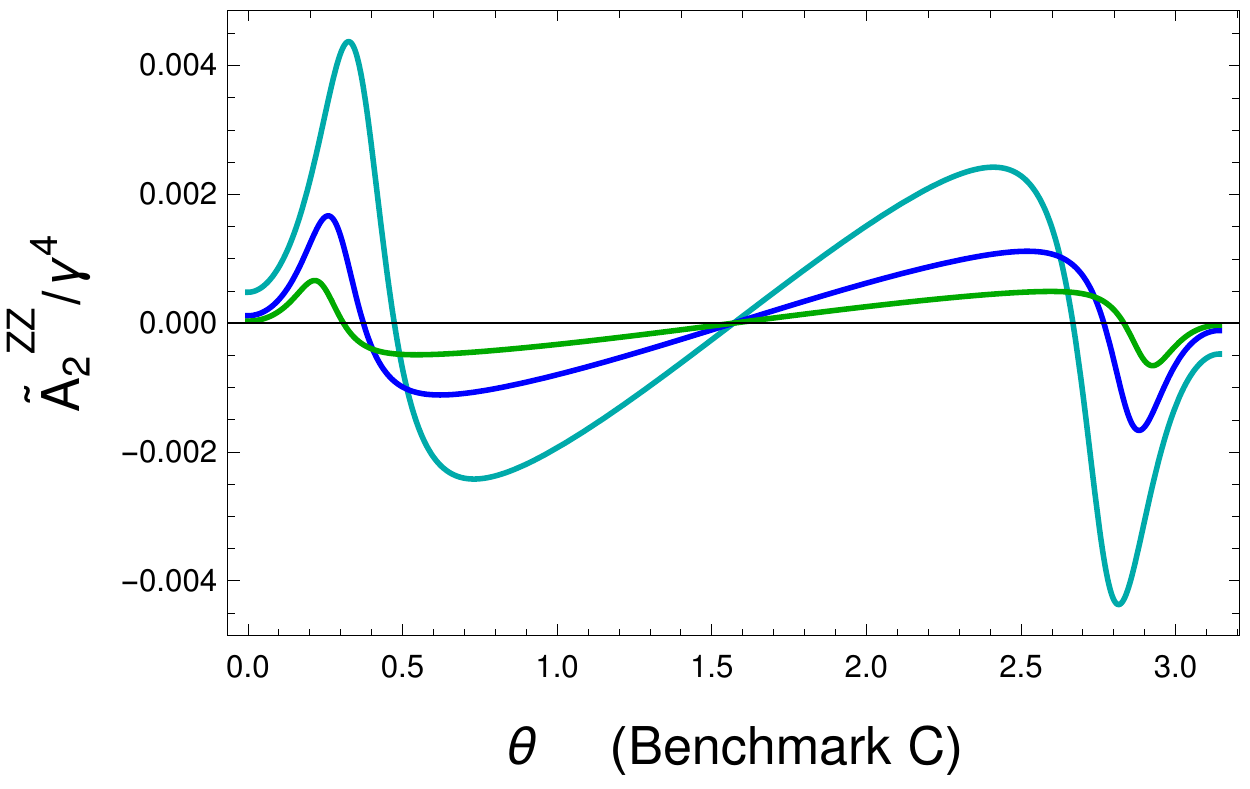}\\
\includegraphics[scale=0.54]{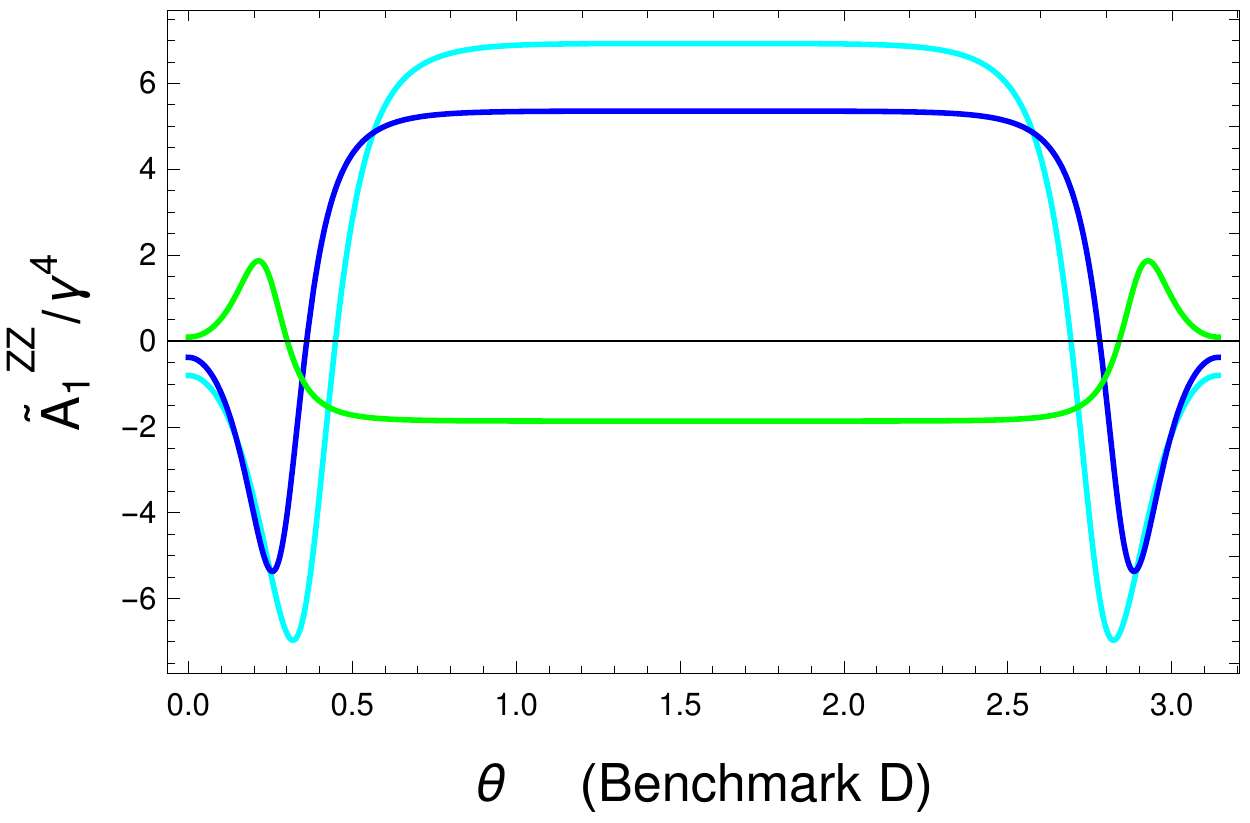}~~~
\includegraphics[scale=0.54]{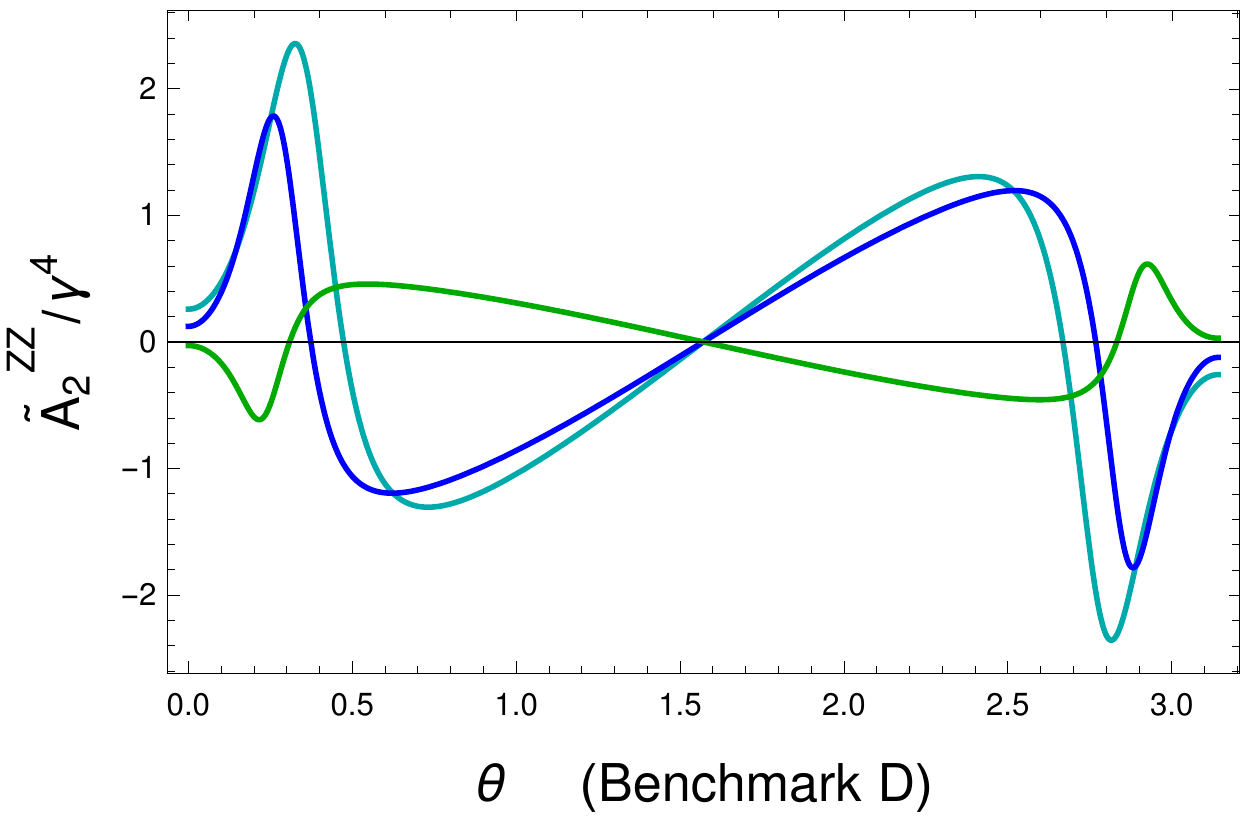}
\caption{The asymmetries $\widetilde{A}^{ZZ}_1(\Theta)$ and $\widetilde{A}^{ZZ}_2(\Theta)$ as functions of $\Theta$ for three beam energies $E$ as indicated (in GeV).}
\label{AAtilde-fig}
\end{center}
\end{figure}

In Figure \ref{AAtilde-fig}, we present the $\widetilde{A}^{ZZ}_1$ and $\widetilde{A}^{ZZ}_2$ asymmetries for all our BPs.

\clearpage

\subsection{Asymmetries ${A^{\prime\prime}_1}^{ZZ}$ and ${A^{\prime\prime}_2}^{ZZ}$ }
To study the helicity formalism of the $Z$ boson pair, it is sufficient to focus on the decay of one outgoing $Z$ boson and study its density matrix, without analysing the complicated event topology of the 4-fermion final state from the decays of the $Z$ boson pairs \cite{Chang:1994cs}. 

The hermitian spin-density matrix $\rho_{\eta,\bar\eta}$ of the $Z$ boson with the scattering angle $\Theta$ (the recoiling $Z$ boson is produced at the scattering angle $\pi - \Theta$) defines the angular distribution of $f'$ in the $Z \to f' \bar{f'}$ decay:
\bea
\rho(\Theta)_{\eta,\bar\eta} \, = \, 
\frac{1}{{\cal N}(\Theta)}
\sum_{\delta,\bar\delta,\,\eta\prime}{\cal M}^{\delta,\bar\delta}_{\eta,\,\eta\prime}\,(\Theta) \,
{{\cal M}^\ast}^{\delta,\bar\delta}_{\bar\eta,\,\eta\prime}\,(\Theta),
\eea
where, again, $\delta, \bar\delta$ are the helicities of  the incoming $f,\bar f$ beams and $\eta, \bar\eta$ are those of the outgoing $Z$ bosons. Here, $\mathcal{N}$ is a normalisation factor which ensures Tr$(\rho)=1$.

\begin{figure}[h!]
\begin{center}
\includegraphics[scale=0.54]{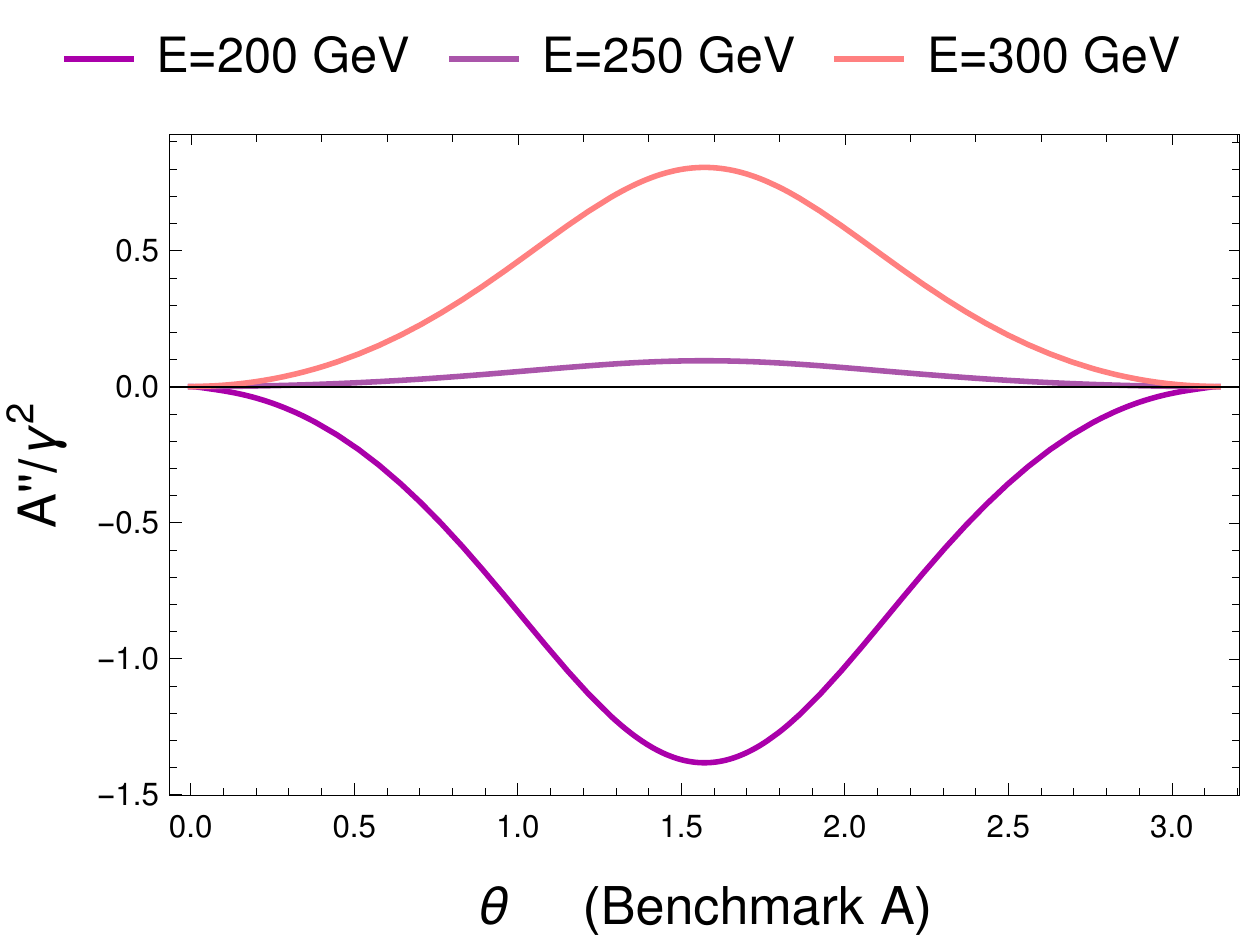}~~~
\includegraphics[scale=0.54]{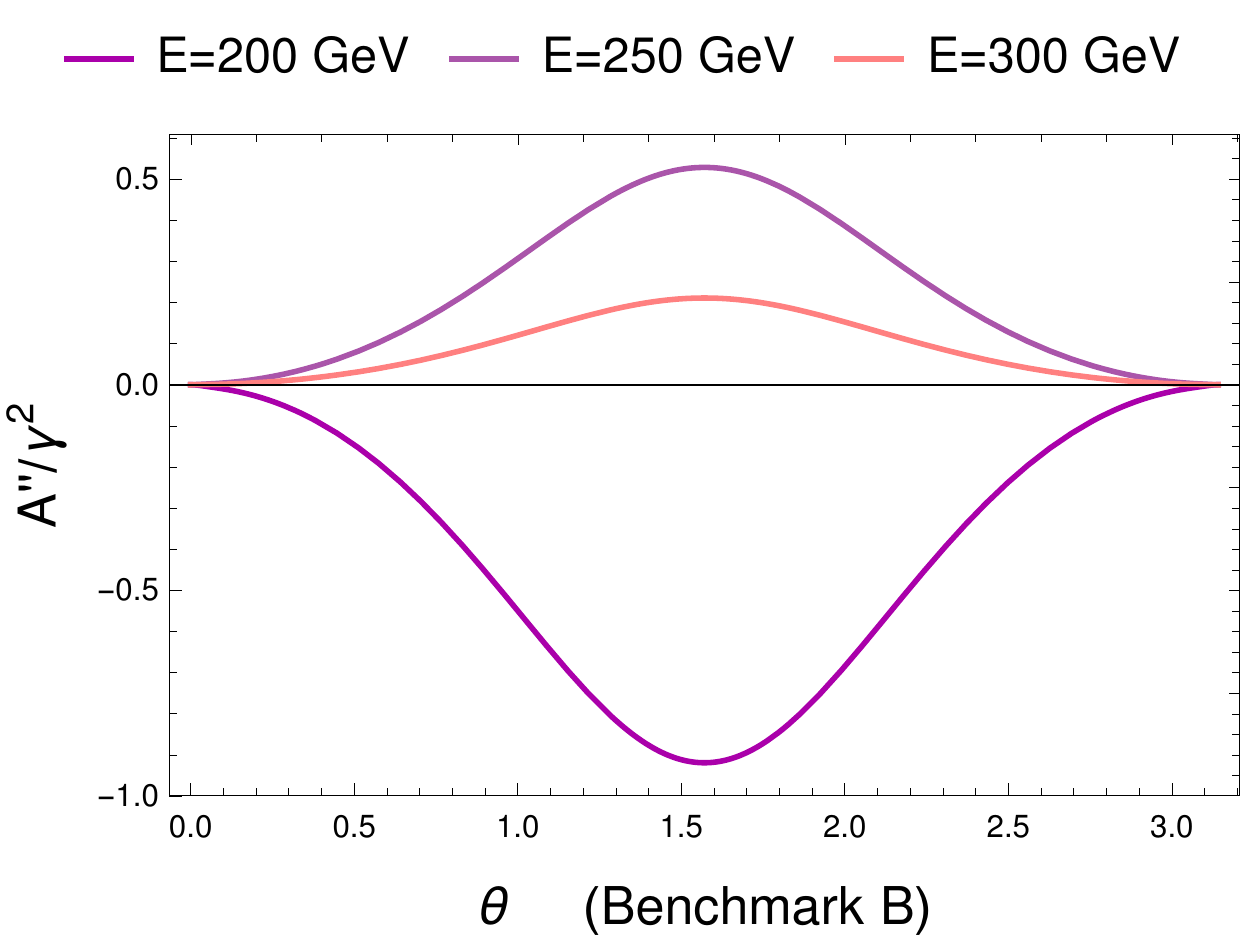}\\
\includegraphics[scale=0.54]{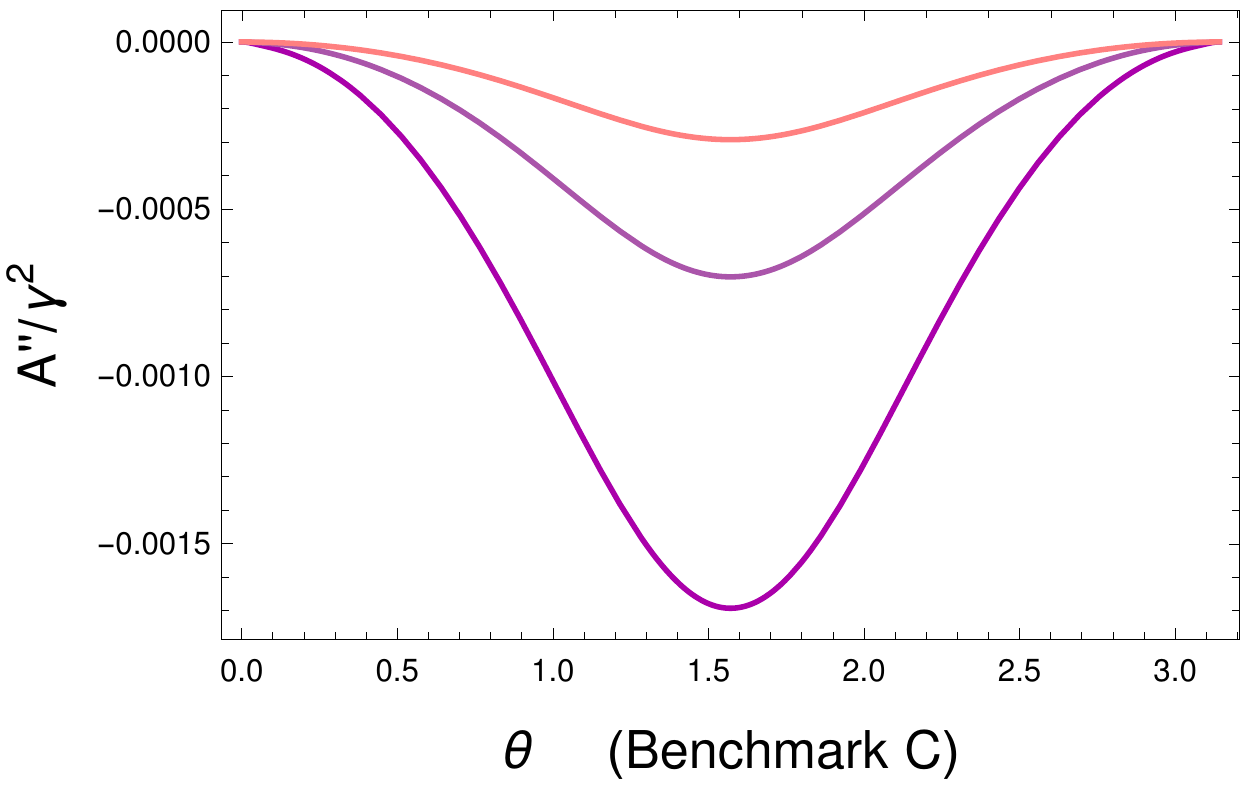}~~~
\includegraphics[scale=0.54]{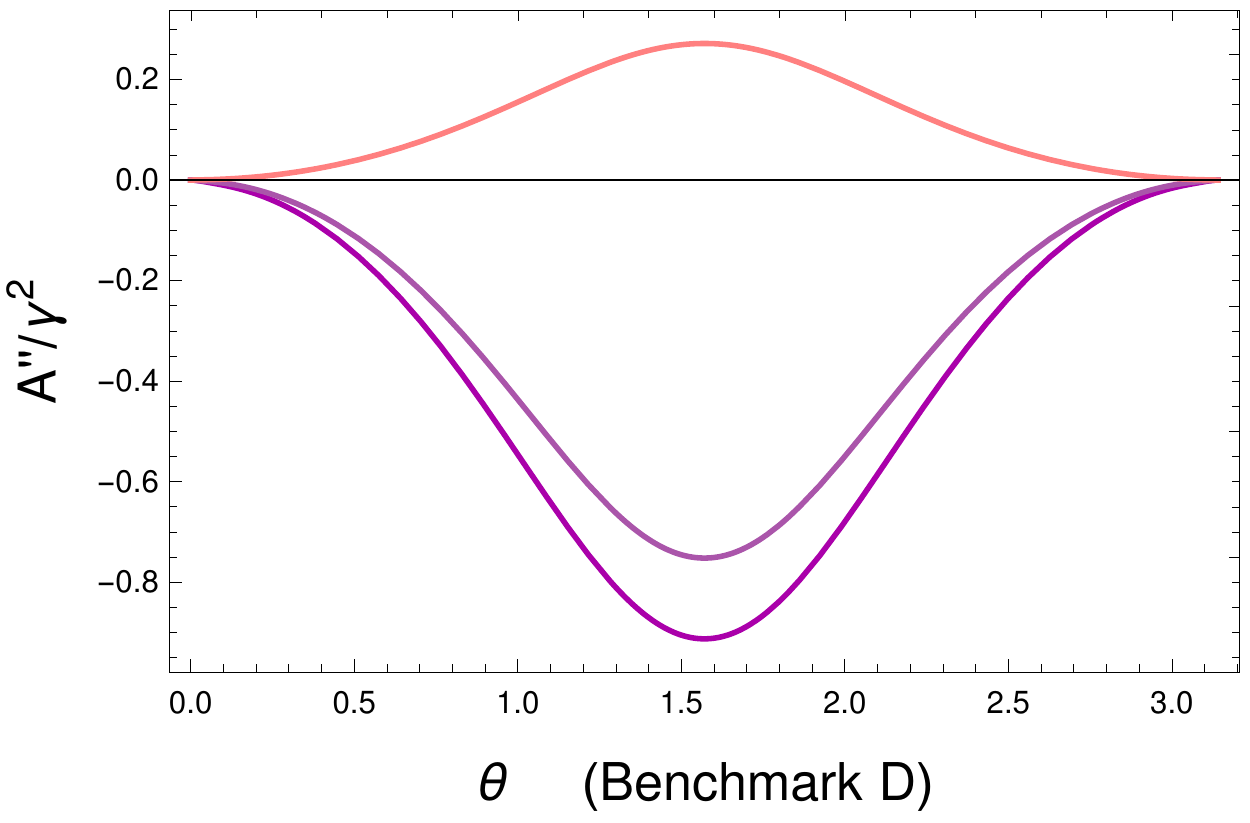}
\caption{The asymmetry ${\cal A}^{\prime\prime}(\Theta)$ as a function of $\Theta$ for three beam energies $E$ as indicated (in GeV).}
\label{App-figures}
\end{center}
\end{figure}

Since the $(+,-)$ or $(-,+)$ components of the spin-density matrix $\rho$ receive the largest CP-violating contribution \cite{Chang:1994cs}, another observable CP-violating asymmetry is defined as 
\be 
A^{\prime\prime}_1
=-\frac{1}{\pi} [
\Im\rho(\Theta)_{+,-} ],
\qquad 
A^{\prime\prime}_2
=\frac{1}{\pi}
[\Im\rho(\pi-\Theta)_{-,+}].
\ee
Calculating this to the lowest order in $f_4^Z$, one finds:
\bea 
\label{App-asymmetry}
{\cal A}^{\prime\prime}(\Theta) &=&
A^{\prime\prime}_1 -
A^{\prime\prime}_2
\\[1mm]
&=& 
\biggl[
\frac{\beta(1+\beta^2)\gamma^2
[(1+\beta^2)^2-(2\beta\cos\Theta)^2]\sin^2\Theta}
{\pi[2+3\beta^2-\beta^6
-\beta^2(9-10\beta^2+\beta^4)\cos^2\Theta
-4\beta^4\cos^4\Theta]}
\biggr]
\, \xi\, \Re f_4^Z,
\nonumber
\eea
\vspace{2mm}
which, unlike other asymmetries defined here, is proportional to the real part of $f_4^Z$.
Figure \ref{App-figures} shows the ${\cal A}^{\prime\prime}(\Theta)$ asymmetry for all our BPs.

\section{Conclusions and outlook}
\label{conclusion}

In this paper, we have shown that CP-violation originating in the {inert} sector of the I(2+1)HDM can make itself manifest in the active one, in fact, in gauge interactions, through one-loop effects entering the cross section for $f\bar f\to Z^*\to ZZ$ at the LHC (and future lepton colliders). This process is mediated by neutral Higgs boson triangle topologies triggered by the inert states of the aforementioned framework.
Unlike the case of the CP-violating 2HDM, where such effects also exist but are limited in size since one of the three neutral states has to be very SM-like, in the I(2+1)HDM all four contributing neutral scalars are inert and can have large gauge couplings. Further, none of the interactions that are generated by the latter can be constrained by EDM data, so that they can all contribute coherently to generate significant asymmetries and increase the cross section for the $f\bar f\to Z^*\to ZZ$ process, above and beyond the 
CP-violating 2HDM yield or that of  the 2HDM plus a singlet.
The 2HDM plus a singlet case with one active doublet scalar and an inert singlet plus doublet scalars not only has fewer number of inert states contributing to the $ZZZ$ loop, but also has diluted $ZS_iS_j$ couplings since the singlet has no direct couplings to the SM gauge bosons.

In order to illustrate such a phenomenology, we have defined several BPs, each embedding CP-violation, over the 
I(2+1)HDM parameter space, with varying mass splittings and coupling strengths in the inert sector, all compliant with available experimental data, from relic density, (in)direct DM searches and colliders. For three such BPs, we have quantified CP-violation effects entering three asymmetries which can all be defined in the $q\bar q\to Z^*\to ZZ$ channel and measured at both the LHC ($f=q$)   by the end of its lifetime  (i.e., after the High Luminosity LHC \cite{Gianotti:2002xx,Abada:2019ono} runs)  and at future lepton colliders ($f=e$) such as the FCC-ee, ILC, CLiC or CEPC running at current design luminosities. Finally, we have illustrated that the hadronic cross sections are typically larger than the leptonic ones, so that it is quite possible that a first evidence of a CP-violating I(2+1)HDM will occur at the LHC rather than at the  FCC-ee, ILC, CLiC or CEPC.

\section*{Acknowledgements}
SM acknowledges support from the STFC Consolidated grant ST/L000296/1 and is financed in part through the NExT Institute. SM, VK, and DR-C acknowledge the H2020-MSCA-RISE-2014 grant no. 645722 (NonMinimalHiggs).
DS is supported in part by the National Science Center, Poland, through the HARMONIA project under contract UMO-2015/18/M/ST2/00518.
DR-C is supported by the Royal Society Newton International Fellowship NIF/R1/180813.
JH-S, DR-C and AC are supported by CONACYT (M\'exico), VIEP-BUAP and 
PRODEP-SEP (M\'exico) under the grant: ``Red Tem\'atica: F\'{\i}sica del Higgs y del Sabor".
VK acknowledges financial support from Academy of Finland projects ``Particle cosmology and gravitational waves'' no. 320123
and ``Particle cosmology beyond the Standard Model'' no. 310130.


\begin{thebibliography}{99}
%\cite{Aad:2012tfa}
\bibitem{Aad:2012tfa} 
  G.~Aad {\it et al.} [ATLAS Collaboration],
  %``Observation of a new particle in the search for the Standard Model Higgs boson with the ATLAS detector at the LHC,''
  Phys.\ Lett.\ B {\bf 716}, 1 (2012)
%  doi:10.1016/j.physletb.2012.08.020
  [arXiv:1207.7214 [hep-ex]].
  %%CITATION = doi:10.1016/j.physletb.2012.08.020;%%
  %10444 citations counted in INSPIRE as of 29 Jan 2020


%\cite{Chatrchyan:2012ufa}
\bibitem{Chatrchyan:2012ufa} 
  S.~Chatrchyan {\it et al.} [CMS Collaboration],
  %``Observation of a New Boson at a Mass of 125 GeV with the CMS Experiment at the LHC,''
  Phys.\ Lett.\ B {\bf 716}, 30 (2012)
%  doi:10.1016/j.physletb.2012.08.021
  [arXiv:1207.7235 [hep-ex]].
  %%CITATION = doi:10.1016/j.physletb.2012.08.021;%%
  %10203 citations counted in INSPIRE as of 29 Jan 2020


%\cite{Aghanim:2018eyx}
\bibitem{Aghanim:2018eyx} 
  N.~Aghanim {\it et al.} [Planck Collaboration],
  %``Planck 2018 results. VI. Cosmological parameters,''
  arXiv:1807.06209 [astro-ph.CO].
  %%CITATION = ARXIV:1807.06209;%%
  %2181 citations counted in INSPIRE as of 29 Jan 2020



%\cite{Jungman:1995df}
\bibitem{Jungman:1995df} 
  G.~Jungman, M.~Kamionkowski and K.~Griest,
  %``Supersymmetric dark matter,''
  Phys.\ Rept.\  {\bf 267}, 195 (1996)
%  doi:10.1016/0370-1573(95)00058-5
  [hep-ph/9506380].
  %%CITATION = doi:10.1016/0370-1573(95)00058-5;%%
  %3911 citations counted in INSPIRE as of 29 Jan 2020


%\cite{Bertone:2004pz}
\bibitem{Bertone:2004pz} 
  G.~Bertone, D.~Hooper and J.~Silk,
  %``Particle dark matter: Evidence, candidates and constraints,''
  Phys.\ Rept.\  {\bf 405}, 279 (2005)
%  doi:10.1016/j.physrep.2004.08.031
  [hep-ph/0404175].
  %%CITATION = doi:10.1016/j.physrep.2004.08.031;%%
  %3603 citations counted in INSPIRE as of 29 Jan 2020


%\cite{Bergstrom:2000pn}
\bibitem{Bergstrom:2000pn} 
  L.~Bergstrom,
  %``Nonbaryonic dark matter: Observational evidence and detection methods,''
  Rept.\ Prog.\ Phys.\  {\bf 63}, 793 (2000)
%  doi:10.1088/0034-4885/63/5/2r3
  [hep-ph/0002126].
  %%CITATION = doi:10.1088/0034-4885/63/5/2r3;%%
  %772 citations counted in INSPIRE as of 29 Jan 2020


%\cite{Deshpande:1977rw}
\bibitem{Deshpande:1977rw} 
  N.~G.~Deshpande and E.~Ma,
  %``Pattern of Symmetry Breaking with Two Higgs Doublets,''
  Phys.\ Rev.\ D {\bf 18}, 2574 (1978).
%  doi:10.1103/PhysRevD.18.2574
  %%CITATION = doi:10.1103/PhysRevD.18.2574;%%
  %695 citations counted in INSPIRE as of 29 Jan 2020


%\cite{Ilnicka:2015jba}
\bibitem{Ilnicka:2015jba} 
  A.~Ilnicka, M.~Krawczyk and T.~Robens,
  %``Inert Doublet Model in light of LHC Run I and astrophysical data,''
  Phys.\ Rev.\ D {\bf 93}, no. 5, 055026 (2016)
%  doi:10.1103/PhysRevD.93.055026
  [arXiv:1508.01671 [hep-ph]].
  %%CITATION = doi:10.1103/PhysRevD.93.055026;%%
  %85 citations counted in INSPIRE as of 29 Jan 2020


%\cite{Belyaev:2016lok}
\bibitem{Belyaev:2016lok} 
  A.~Belyaev, G.~Cacciapaglia, I.~P.~Ivanov, F.~Rojas-Abatte and M.~Thomas,
  %``Anatomy of the Inert Two Higgs Doublet Model in the light of the LHC and non-LHC Dark Matter Searches,''
  Phys.\ Rev.\ D {\bf 97}, no. 3, 035011 (2018)
%  doi:10.1103/PhysRevD.97.035011
  [arXiv:1612.00511 [hep-ph]].
  %%CITATION = doi:10.1103/PhysRevD.97.035011;%%
  %68 citations counted in INSPIRE as of 29 Jan 2020


%\cite{Belyaev:2018ext}
\bibitem{Belyaev:2018ext} 
  A.~Belyaev {\it et al.},
  %``Advancing LHC probes of dark matter from the inert two-Higgs-doublet model with the monojet signal,''
  Phys.\ Rev.\ D {\bf 99}, no. 1, 015011 (2019)
%  doi:10.1103/PhysRevD.99.015011
  [arXiv:1809.00933 [hep-ph]].
  %%CITATION = doi:10.1103/PhysRevD.99.015011;%%
  %18 citations counted in INSPIRE as of 29 Jan 2020


%\cite{Kalinowski:2018ylg}
\bibitem{Kalinowski:2018ylg} 
  J.~Kalinowski, W.~Kotlarski, T.~Robens, D.~Sokolowska and A.~F.~Zarnecki,
  %``Benchmarking the Inert Doublet Model for $e^+ e^-$ colliders,''
  JHEP {\bf 1812}, 081 (2018)
%  doi:10.1007/JHEP12(2018)081
  [arXiv:1809.07712 [hep-ph]].
  %%CITATION = doi:10.1007/JHEP12(2018)081;%%
  %19 citations counted in INSPIRE as of 29 Jan 2020



%\cite{Keus:2014jha}
\bibitem{Keus:2014jha} 
  V.~Keus, S.~F.~King, S.~Moretti and D.~Sokolowska,
  %``Dark Matter with Two Inert Doublets plus One Higgs Doublet,''
  JHEP {\bf 1411}, 016 (2014)
%  doi:10.1007/JHEP11(2014)016
  [arXiv:1407.7859 [hep-ph]].
  %%CITATION = doi:10.1007/JHEP11(2014)016;%%
  %37 citations counted in INSPIRE as of 29 Jan 2020


%\cite{Keus:2015xya}
\bibitem{Keus:2015xya} 
  V.~Keus, S.~F.~King, S.~Moretti and D.~Sokolowska,
  %``Observable Heavy Higgs Dark Matter,''
  JHEP {\bf 1511}, 003 (2015)
%  doi:10.1007/JHEP11(2015)003
  [arXiv:1507.08433 [hep-ph]].
  %%CITATION = doi:10.1007/JHEP11(2015)003;%%
  %16 citations counted in INSPIRE as of 29 Jan 2020


%\cite{Keus:2013hya}
\bibitem{Keus:2013hya} 
  V.~Keus, S.~F.~King and S.~Moretti,
  %``Three-Higgs-doublet models: symmetries, potentials and Higgs boson masses,''
  JHEP {\bf 1401}, 052 (2014)
%  doi:10.1007/JHEP01(2014)052
  [arXiv:1310.8253 [hep-ph]].
  %%CITATION = doi:10.1007/JHEP01(2014)052;%%
  %49 citations counted in INSPIRE as of 29 Jan 2020


%\cite{Cordero-Cid:2016krd}
\bibitem{Cordero-Cid:2016krd} 
  A.~Cordero-Cid, J.~Hernandez-Sanchez, V.~Keus, S.~F.~King, S.~Moretti, D.~Rojas and D.~Sokolowska,
  %``CP violating scalar Dark Matter,''
  JHEP {\bf 1612}, 014 (2016)
%  doi:10.1007/JHEP12(2016)014
  [arXiv:1608.01673 [hep-ph]].
  %%CITATION = doi:10.1007/JHEP12(2016)014;%%
  %12 citations counted in INSPIRE as of 29 Jan 2020


%\cite{Keus:2019szx}
\bibitem{Keus:2019szx} 
  V.~Keus,
  %``Dark CP-violation through the $Z$-portal,''
  arXiv:1909.09234 [hep-ph].
  %%CITATION = ARXIV:1909.09234;%%
  %1 citations counted in INSPIRE as of 29 Jan 2020

%\cite{Cordero-Cid:2018man}
\bibitem{Cordero-Cid:2018man} 
  A.~Cordero-Cid, J.~Hernandez-Sanchez, V.~Keus, S.~Moretti, D.~Rojas and D.~Sokolowska,
  %``Lepton collider signatures of dark CP violation,''
  arXiv:1812.00820 [hep-ph].
  %%CITATION = ARXIV:1812.00820;%%
  %2 citations counted in INSPIRE as of 29 Jan 2020

%\cite{Grzadkowski:2009bt}
\bibitem{Grzadkowski:2009bt} 
  B.~Grzadkowski, O.~M.~Ogreid and P.~Osland,
  %``Natural Multi-Higgs Model with Dark Matter and CP Violation,''
  Phys.\ Rev.\ D {\bf 80}, 055013 (2009)
%  doi:10.1103/PhysRevD.80.055013
  [arXiv:0904.2173 [hep-ph]].
  %%CITATION = doi:10.1103/PhysRevD.80.055013;%%
  %35 citations counted in INSPIRE as of 29 Jan 2020


%\cite{Osland:2013sla}
\bibitem{Osland:2013sla} 
  P.~Osland, A.~Pukhov, G.~M.~Pruna and M.~Purmohammadi,
  %``Phenomenology of charged scalars in the CP-Violating Inert-Doublet Model,''
  JHEP {\bf 1304}, 040 (2013)
%  doi:10.1007/JHEP04(2013)040
  [arXiv:1302.3713 [hep-ph]].
  %%CITATION = doi:10.1007/JHEP04(2013)040;%%
  %13 citations counted in INSPIRE as of 29 Jan 2020

%\cite{Keus:2015hva}
\bibitem{Keus:2015hva} 
  V.~Keus, S.~F.~King, S.~Moretti and K.~Yagyu,
  %``CP Violating Two-Higgs-Doublet Model: Constraints and LHC Predictions,''
  JHEP {\bf 1604}, 048 (2016)
%  doi:10.1007/JHEP04(2016)048
  [arXiv:1510.04028 [hep-ph]].
  %%CITATION = doi:10.1007/JHEP04(2016)048;%%
  %17 citations counted in INSPIRE as of 10 Feb 2020

%\cite{Keus:2017ioh}
\bibitem{Keus:2017ioh} 
  V.~Keus, N.~Koivunen and K.~Tuominen,
  %``Singlet scalar and 2HDM extensions of the Standard Model: CP-violation and constraints from $(g-2)_\mu$ and $e$EDM,''
  JHEP {\bf 1809}, 059 (2018)
%  doi:10.1007/JHEP09(2018)059
  [arXiv:1712.09613 [hep-ph]].
  %%CITATION = doi:10.1007/JHEP09(2018)059;%%
  %6 citations counted in INSPIRE as of 10 Feb 2020


%\cite{Ivanov:2011ae}
\bibitem{Ivanov:2011ae} 
  I.~P.~Ivanov, V.~Keus and E.~Vdovin,
  %``Abelian symmetries in multi-Higgs-doublet models,''
  J.\ Phys.\ A {\bf 45}, 215201 (2012)
%  doi:10.1088/1751-8113/45/21/215201
  [arXiv:1112.1660 [math-ph]].
  %%CITATION = doi:10.1088/1751-8113/45/21/215201;%%
  %54 citations counted in INSPIRE as of 10 Feb 2020



%\cite{Sirunyan:2019twz}
\bibitem{Sirunyan:2019twz} 
  A.~M.~Sirunyan {\it et al.} [CMS Collaboration],
  %``Measurements of the Higgs boson width and anomalous $HVV$ couplings from on-shell and off-shell production in the four-lepton final state,''
  Phys.\ Rev.\ D {\bf 99}, no. 11, 112003 (2019)
%  doi:10.1103/PhysRevD.99.112003
  [arXiv:1901.00174 [hep-ex]].
  %%CITATION = doi:10.1103/PhysRevD.99.112003;%%
  %39 citations counted in INSPIRE as of 29 Jan 2020


%\cite{Aaboud:2018xdt}
\bibitem{Aaboud:2018xdt} 
  M.~Aaboud {\it et al.} [ATLAS Collaboration],
  %``Measurements of Higgs boson properties in the diphoton decay channel with 36 fb$^{-1}$ of $pp$ collision data at $\sqrt{s} = 13$ TeV with the ATLAS detector,''
  Phys.\ Rev.\ D {\bf 98}, 052005 (2018)
%  doi:10.1103/PhysRevD.98.052005
  [arXiv:1802.04146 [hep-ex]].
  %%CITATION = doi:10.1103/PhysRevD.98.052005;%%
  %152 citations counted in INSPIRE as of 29 Jan 2020


%\cite{Sirunyan:2018ouh}
\bibitem{Sirunyan:2018ouh} 
  A.~M.~Sirunyan {\it et al.} [CMS Collaboration],
  %``Measurements of Higgs boson properties in the diphoton decay channel in proton-proton collisions at $\sqrt{s} =$ 13 TeV,''
  JHEP {\bf 1811}, 185 (2018)
%  doi:10.1007/JHEP11(2018)185
  [arXiv:1804.02716 [hep-ex]].
  %%CITATION = doi:10.1007/JHEP11(2018)185;%%
  %107 citations counted in INSPIRE as of 29 Jan 2020


%\cite{Altarelli:1990zd}
\bibitem{Altarelli:1990zd} 
  G.~Altarelli and R.~Barbieri,
  %``Vacuum polarization effects of new physics on electroweak processes,''
  Phys.\ Lett.\ B {\bf 253}, 161 (1991).
%  doi:10.1016/0370-2693(91)91378-9
  %%CITATION = doi:10.1016/0370-2693(91)91378-9;%%
  %803 citations counted in INSPIRE as of 29 Jan 2020


%\cite{Peskin:1990zt}
\bibitem{Peskin:1990zt} 
  M.~E.~Peskin and T.~Takeuchi,
  %``A New constraint on a strongly interacting Higgs sector,''
  Phys.\ Rev.\ Lett.\  {\bf 65}, 964 (1990).
%  doi:10.1103/PhysRevLett.65.964
  %%CITATION = doi:10.1103/PhysRevLett.65.964;%%
  %1877 citations counted in INSPIRE as of 29 Jan 2020


%\cite{Peskin:1991sw}
\bibitem{Peskin:1991sw} 
  M.~E.~Peskin and T.~Takeuchi,
  %``Estimation of oblique electroweak corrections,''
  Phys.\ Rev.\ D {\bf 46}, 381 (1992).
%  doi:10.1103/PhysRevD.46.381
  %%CITATION = doi:10.1103/PhysRevD.46.381;%%
  %2160 citations counted in INSPIRE as of 29 Jan 2020


%\cite{Maksymyk:1993zm}
\bibitem{Maksymyk:1993zm} 
  I.~Maksymyk, C.~P.~Burgess and D.~London,
  %``Beyond S, T and U,''
  Phys.\ Rev.\ D {\bf 50}, 529 (1994)
%  doi:10.1103/PhysRevD.50.529
  [hep-ph/9306267].
  %%CITATION = doi:10.1103/PhysRevD.50.529;%%
  %196 citations counted in INSPIRE as of 29 Jan 2020


%\cite{Pierce:2007ut}
\bibitem{Pierce:2007ut} 
  A.~Pierce and J.~Thaler,
  %``Natural Dark Matter from an Unnatural Higgs Boson and New Colored Particles at the TeV Scale,''
  JHEP {\bf 0708}, 026 (2007)
%  doi:10.1088/1126-6708/2007/08/026
  [hep-ph/0703056 [HEP-PH]].
  %%CITATION = doi:10.1088/1126-6708/2007/08/026;%%
  %123 citations counted in INSPIRE as of 29 Jan 2020


%\cite{Heisig:2018kfq}
\bibitem{Heisig:2018kfq} 
  J.~Heisig, S.~Kraml and A.~Lessa,
  %``Constraining new physics with searches for long-lived particles: Implementation into SModelS,''
  Phys.\ Lett.\ B {\bf 788}, 87 (2019)
%  doi:10.1016/j.physletb.2018.10.049
  [arXiv:1808.05229 [hep-ph]].
  %%CITATION = doi:10.1016/j.physletb.2018.10.049;%%
  %13 citations counted in INSPIRE as of 29 Jan 2020


%\cite{Lundstrom:2008ai}
\bibitem{Lundstrom:2008ai} 
  E.~Lundstrom, M.~Gustafsson and J.~Edsjo,
  %``The Inert Doublet Model and LEP II Limits,''
  Phys.\ Rev.\ D {\bf 79}, 035013 (2009)
%  doi:10.1103/PhysRevD.79.035013
  [arXiv:0810.3924 [hep-ph]].
  %%CITATION = doi:10.1103/PhysRevD.79.035013;%%
  %191 citations counted in INSPIRE as of 29 Jan 2020


%\cite{Aprile:2018dbl}
\bibitem{Aprile:2018dbl} 
  E.~Aprile {\it et al.} [XENON Collaboration],
  %``Dark Matter Search Results from a One Ton-Year Exposure of XENON1T,''
  Phys.\ Rev.\ Lett.\  {\bf 121}, no. 11, 111302 (2018)
%  doi:10.1103/PhysRevLett.121.111302
  [arXiv:1805.12562 [astro-ph.CO]].
  %%CITATION = doi:10.1103/PhysRevLett.121.111302;%%
  %620 citations counted in INSPIRE as of 29 Jan 2020



%\cite{Lavoura:1994fv}
\bibitem{Lavoura:1994fv} 
  L.~Lavoura and J.~P.~Silva,
  %``Fundamental CP violating quantities in a SU(2) x U(1) model with many Higgs doublets,''
  Phys.\ Rev.\ D {\bf 50}, 4619 (1994)
%  doi:10.1103/PhysRevD.50.4619
  [hep-ph/9404276].
  %%CITATION = doi:10.1103/PhysRevD.50.4619;%%
  %161 citations counted in INSPIRE as of 29 Jan 2020


%\cite{Botella:1994cs}
\bibitem{Botella:1994cs} 
  F.~J.~Botella and J.~P.~Silva,
  %``Jarlskog - like invariants for theories with scalars and fermions,''
  Phys.\ Rev.\ D {\bf 51}, 3870 (1995)
%  doi:10.1103/PhysRevD.51.3870
  [hep-ph/9411288].
  %%CITATION = doi:10.1103/PhysRevD.51.3870;%%
  %172 citations counted in INSPIRE as of 29 Jan 2020


%\cite{Gunion:2005ja}
\bibitem{Gunion:2005ja} 
  J.~F.~Gunion and H.~E.~Haber,
  %``Conditions for CP-violation in the general two-Higgs-doublet model,''
  Phys.\ Rev.\ D {\bf 72}, 095002 (2005)
%  doi:10.1103/PhysRevD.72.095002
  [hep-ph/0506227].
  %%CITATION = doi:10.1103/PhysRevD.72.095002;%%
  %171 citations counted in INSPIRE as of 29 Jan 2020


%\cite{Grzadkowski:2014ada}
\bibitem{Grzadkowski:2014ada} 
  B.~Grzadkowski, O.~M.~Ogreid and P.~Osland,
  %``Measuring CP violation in Two-Higgs-Doublet models in light of the LHC Higgs data,''
  JHEP {\bf 1411}, 084 (2014)
%  doi:10.1007/JHEP11(2014)084
  [arXiv:1409.7265 [hep-ph]].
  %%CITATION = doi:10.1007/JHEP11(2014)084;%%
  %32 citations counted in INSPIRE as of 29 Jan 2020


%\cite{Haber:2006ue}
\bibitem{Haber:2006ue} 
  H.~E.~Haber and D.~O'Neil,
  %``Basis-independent methods for the two-Higgs-doublet model. II. The Significance of tan$\beta$,''
  Phys.\ Rev.\ D {\bf 74}, 015018 (2006)
  [Erratum: Phys.\ Rev.\ D {\bf 74}, no. 5, 059905 (2006)]
%  doi:10.1103/PhysRevD.74.015018, 10.1103/PhysRevD.74.059905
  [hep-ph/0602242].
  %%CITATION = doi:10.1103/PhysRevD.74.015018, 10.1103/PhysRevD.74.059905;%%
  %155 citations counted in INSPIRE as of 29 Jan 2020


%\cite{Haber:2015pua}
\bibitem{Haber:2015pua} 
  H.~E.~Haber and O.~Stal,
  %``New LHC benchmarks for the $\mathcal{CP}$ -conserving two-Higgs-doublet model,''
  Eur.\ Phys.\ J.\ C {\bf 75}, no. 10, 491 (2015)
  [Erratum: Eur.\ Phys.\ J.\ C {\bf 76}, no. 6, 312 (2016)]
%  doi:10.1140/epjc/s10052-015-3697-x, 10.1140/epjc/s10052-016-4151-4
  [arXiv:1507.04281 [hep-ph]].
  %%CITATION = doi:10.1140/epjc/s10052-015-3697-x, 10.1140/epjc/s10052-016-4151-4;%%
  %85 citations counted in INSPIRE as of 29 Jan 2020


%\cite{He:1992qh}
\bibitem{He:1992qh} 
  X.~G.~He, J.~P.~Ma and B.~H.~J.~McKellar,
  %``CP violating form-factors for three gauge boson vertex in the two Higgs doublet and left-right symmetric models,''
  Phys.\ Lett.\ B {\bf 304}, 285 (1993)
%  doi:10.1016/0370-2693(93)90296-T
  [hep-ph/9209260].
  %%CITATION = doi:10.1016/0370-2693(93)90296-T;%%
  %18 citations counted in INSPIRE as of 29 Jan 2020


%\cite{Chang:1994cs}
\bibitem{Chang:1994cs} 
  D.~Chang, W.~Y.~Keung and P.~B.~Pal,
  %``CP violation in the cubic coupling of neutral gauge bosons,''
  Phys.\ Rev.\ D {\bf 51}, 1326 (1995)
%  doi:10.1103/PhysRevD.51.1326
  [hep-ph/9407294].
  %%CITATION = doi:10.1103/PhysRevD.51.1326;%%
  %33 citations counted in INSPIRE as of 29 Jan 2020


%\cite{Chang:1993vv}
\bibitem{Chang:1993vv} 
  D.~Chang, W.~Y.~Keung and I.~Phillips,
  %``CP violating observables in e- e+ ---> W- W+,''
  Phys.\ Rev.\ D {\bf 48}, 4045 (1993)
% doi:10.1103/PhysRevD.48.4045
  [hep-ph/9307232].
  %%CITATION = doi:10.1103/PhysRevD.48.4045;%%
  %21 citations counted in INSPIRE as of 29 Jan 2020


%\cite{Hagiwara:1986vm}
\bibitem{Hagiwara:1986vm} 
  K.~Hagiwara, R.~D.~Peccei, D.~Zeppenfeld and K.~Hikasa,
  %``Probing the Weak Boson Sector in e+ e- ---> W+ W-,''
  Nucl.\ Phys.\ B {\bf 282}, 253 (1987).
%  doi:10.1016/0550-3213(87)90685-7
  %%CITATION = doi:10.1016/0550-3213(87)90685-7;%%
  %1379 citations counted in INSPIRE as of 29 Jan 2020


%\cite{Gounaris:1999kf}
\bibitem{Gounaris:1999kf} 
  G.~J.~Gounaris, J.~Layssac and F.~M.~Renard,
  %``Signatures of the anomalous $Z_{\gamma}$ and $Z Z$ production at the lepton and hadron colliders,''
  Phys.\ Rev.\ D {\bf 61}, 073013 (2000)
%  doi:10.1103/PhysRevD.61.073013
  [hep-ph/9910395].
  %%CITATION = doi:10.1103/PhysRevD.61.073013;%%
  %95 citations counted in INSPIRE as of 29 Jan 2020


%\cite{Gounaris:2000dn}
\bibitem{Gounaris:2000dn} 
  G.~J.~Gounaris, J.~Layssac and F.~M.~Renard,
  %``Off-shell structure of the anomalous $Z$ and $\gamma$ selfcouplings,''
  Phys.\ Rev.\ D {\bf 65}, 017302 (2002)
  [Phys.\ Rev.\ D {\bf 62}, 073012 (2000)]
%  doi:10.1103/PhysRevD.62.073012, 10.1103/PhysRevD.65.017302
  [hep-ph/0005269].
  %%CITATION = doi:10.1103/PhysRevD.62.073012, 10.1103/PhysRevD.65.017302;%%
  %65 citations counted in INSPIRE as of 29 Jan 2020


%\cite{Baur:2000ae}
\bibitem{Baur:2000ae} 
  U.~Baur and D.~L.~Rainwater,
  %``Probing neutral gauge boson selfinteractions in $ZZ$ production at hadron colliders,''
  Phys.\ Rev.\ D {\bf 62}, 113011 (2000)
%  doi:10.1103/PhysRevD.62.113011
  [hep-ph/0008063].
  %%CITATION = doi:10.1103/PhysRevD.62.113011;%%
  %78 citations counted in INSPIRE as of 29 Jan 2020


%\cite{Grzadkowski:2016lpv}
\bibitem{Grzadkowski:2016lpv} 
  B.~Grzadkowski, O.~M.~Ogreid and P.~Osland,
  %``CP-Violation in the $ZZZ$ and $ZWW$ vertices at $e^+e^-$ colliders in Two-Higgs-Doublet Models,''
  JHEP {\bf 1605}, 025 (2016)
  Erratum: [JHEP {\bf 1711}, 002 (2017)]
%  doi:10.1007/JHEP05(2016)025, 10.1007/JHEP11(2017)002
  [arXiv:1603.01388 [hep-ph]].
  %%CITATION = doi:10.1007/JHEP05(2016)025, 10.1007/JHEP11(2017)002;%%
  %15 citations counted in INSPIRE as of 29 Jan 2020


%\cite{Hahn:2010zi}
\bibitem{Hahn:2010zi} 
  T.~Hahn,
  %``Feynman Diagram Calculations with FeynArts, FormCalc, and LoopTools,''
  PoS ACAT {\bf 2010}, 078 (2010)
%  doi:10.22323/1.093.0078
  [arXiv:1006.2231 [hep-ph]].
  %%CITATION = doi:10.22323/1.093.0078;%%
  %56 citations counted in INSPIRE as of 29 Jan 2020


%\cite{Gounaris:2000tb}
\bibitem{Gounaris:2000tb} 
  G.~J.~Gounaris, J.~Layssac and F.~M.~Renard,
  %``New and standard physics contributions to anomalous Z and gamma selfcouplings,''
  Phys.\ Rev.\ D {\bf 62}, 073013 (2000)
%  doi:10.1103/PhysRevD.62.073013
  [hep-ph/0003143].
  %%CITATION = doi:10.1103/PhysRevD.62.073013;%%
  %83 citations counted in INSPIRE as of 29 Jan 2020


%\cite{Moyotl:2015bia}
\bibitem{Moyotl:2015bia} 
  A.~Moyotl, J.~J.~Toscano and G.~Tavares-Velasco,
  %``CP-odd contributions to the $ZZ^\ast\gamma$, $ZZ\gamma^\ast$, and $ZZZ^\ast$ vertices induced by nondiagonal charged scalar boson couplings,''
  Phys.\ Rev.\ D {\bf 91}, 093005 (2015)
%  doi:10.1103/PhysRevD.91.093005
  [arXiv:1505.01253 [hep-ph]].
  %%CITATION = doi:10.1103/PhysRevD.91.093005;%%
  %2 citations counted in INSPIRE as of 29 Jan 2020


%\cite{Azevedo:2018fmj}
\bibitem{Azevedo:2018fmj} 
  D.~Azevedo, P.~M.~Ferreira, M.~M.~Muhlleitner, S.~Patel, R.~Santos and J.~Wittbrodt,
  %``CP in the dark,''
  JHEP {\bf 1811}, 091 (2018)
%  doi:10.1007/JHEP11(2018)091
  [arXiv:1807.10322 [hep-ph]].
  %%CITATION = doi:10.1007/JHEP11(2018)091;%%
  %6 citations counted in INSPIRE as of 29 Jan 2020


%\cite{Belyaev:2012qa}
\bibitem{Belyaev:2012qa} 
  A.~Belyaev, N.~D.~Christensen and A.~Pukhov,
  %``CalcHEP 3.4 for collider physics within and beyond the Standard Model,''
  Comput.\ Phys.\ Commun.\  {\bf 184}, 1729 (2013)
%  doi:10.1016/j.cpc.2013.01.014
  [arXiv:1207.6082 [hep-ph]].
  %%CITATION = doi:10.1016/j.cpc.2013.01.014;%%
  %709 citations counted in INSPIRE as of 29 Jan 2020

%\cite{Stump:2003yu}
\bibitem{Stump:2003yu} 
  D.~Stump, J.~Huston, J.~Pumplin, W.~K.~Tung, H.~L.~Lai, S.~Kuhlmann and J.~F.~Owens,
  %``Inclusive jet production, parton distributions, and the search for new physics,''
  JHEP {\bf 0310}, 046 (2003)
%  doi:10.1088/1126-6708/2003/10/046
  [hep-ph/0303013].
  %%CITATION = doi:10.1088/1126-6708/2003/10/046;%%
  %978 citations counted in INSPIRE as of 10 Feb 2020

%\cite{Craig:2017gzf}
\bibitem{Craig:2017gzf} 
  N.~Craig,
  %``A Case for Future Lepton Colliders,''
  arXiv:1703.06079 [hep-ph].
  %%CITATION = ARXIV:1703.06079;%%
  %12 citations counted in INSPIRE as of 10 Feb 2020


%\cite{Djouadi:2007ik}
\bibitem{Djouadi:2007ik} 
  A.~Djouadi {\it et al.} [ILC Collaboration],
  %``International Linear Collider Reference Design Report Volume 2: Physics at the ILC,''
  arXiv:0709.1893 [hep-ph].
  %%CITATION = ARXIV:0709.1893;%%
  %420 citations counted in INSPIRE as of 29 Jan 2020


%\cite{Lebrun:2012hj}
\bibitem{Lebrun:2012hj} 
  P.~Lebrun {\it et al.},
  %``The CLIC Programme: Towards a Staged e+e- Linear Collider Exploring the Terascale : CLIC Conceptual Design Report,''
%  doi:10.5170/CERN-2012-005
  arXiv:1209.2543 [physics.ins-det].
  %%CITATION = doi:10.5170/CERN-2012-005;%%
  %153 citations counted in INSPIRE as of 29 Jan 2020


%\cite{Gounaris:1991ce}
\bibitem{Gounaris:1991ce} 
  G.~Gounaris, D.~Schildknecht and F.~M.~Renard,
  %``Genuine tests of CP invariance in e+ e- ---> W+ W-,''
  Phys.\ Lett.\ B {\bf 263}, 291 (1991).
%  doi:10.1016/0370-2693(91)90603-N
  %%CITATION = doi:10.1016/0370-2693(91)90603-N;%%
  %56 citations counted in INSPIRE as of 29 Jan 2020

%\cite{Gianotti:2002xx}
\bibitem{Gianotti:2002xx} 
  F.~Gianotti {\it et al.},
  %``Physics potential and experimental challenges of the LHC luminosity upgrade,''
  Eur.\ Phys.\ J.\ C {\bf 39}, 293 (2005)
 % doi:10.1140/epjc/s2004-02061-6
  [hep-ph/0204087].
  %%CITATION = doi:10.1140/epjc/s2004-02061-6;%%
  %422 citations counted in INSPIRE as of 10 Feb 2020
  
%\cite{Abada:2019ono}
\bibitem{Abada:2019ono} 
  A.~Abada {\it et al.} [FCC Collaboration],
  %``HE-LHC: The High-Energy Large Hadron Collider : Future Circular Collider Conceptual Design Report Volume 4,''
  Eur.\ Phys.\ J.\ ST {\bf 228}, no. 5, 1109 (2019).
%  doi:10.1140/epjst/e2019-900088-6
  %%CITATION = doi:10.1140/epjst/e2019-900088-6;%%
  %53 citations counted in INSPIRE as of 10 Feb 2020

    
\end{thebibliography}
\end{document}